\pdfoutput=1
\RequirePackage{etoolbox}
\newbool{fullversion}
\booltrue{fullversion}

\documentclass[acmsmall]{acmart}%

\usepackage{etoolbox}

\providebool{fullversion}

\usepackage{preamble}
\usepackage{iris}
\usepackage{prob}
\usepackage{examples}

\setcopyright{rightsretained}
\acmDOI{10.1145/3689753}
\acmYear{2024}
\acmJournal{PACMPL}
\acmVolume{8}
\acmNumber{OOPSLA2}
\acmArticle{313}
\acmMonth{10}
\acmSubmissionID{oopslab24main-p414-p}
\received{2024-04-05}
\received[accepted]{2024-08-18}



\begin{document}

\title{\thetimelogic: Higher-Order Separation Logic with Credits for Expected Costs}

\author[P. G. Haselwarter]{Philipp~G. Haselwarter}
\orcid{0000-0003-0198-7751}
\affiliation{
  \institution{Aarhus University}
  \country{Denmark}
}
\email{pgh@cs.au.dk}

\author[K. H. Li]{Kwing Hei Li}
\orcid{0000-0002-4124-5720}
\affiliation{
  \institution{Aarhus University}
  \country{Denmark}
}
\email{hei.li@cs.au.dk}

\author[M. de Medeiros]{Markus de Medeiros}
\orcid{0009-0005-3285-5032}
\affiliation{
  \institution{New York University}
  \country{USA}
}
\email{mjd9606@nyu.edu}

\author[S. O. Gregersen]{Simon Oddershede Gregersen}
\orcid{0000-0001-6045-5232}
\affiliation{
  \institution{New York University}
  \country{USA}
}
\email{s.gregersen@nyu.edu}

\author[A. Aguirre]{Alejandro Aguirre}
\orcid{0000-0001-6746-2734}
\affiliation{
  \institution{Aarhus University}
  \country{Denmark}
}
\email{alejandro@cs.au.dk}

\author[J. Tassarotti]{Joseph Tassarotti}
\orcid{0000-0001-5692-3347}
\affiliation{
  \institution{New York University}
  \country{USA}
}
\email{jt4767@nyu.edu}

\author[L. Birkedal]{Lars Birkedal}
\orcid{0000-0003-1320-0098}
\affiliation{
  \institution{Aarhus University}
  \country{Denmark}
}
\email{birkedal@cs.au.dk}

\begin{abstract}
  We present \thetimelogic, a higher-order separation logic to reason about
  the expected cost of probabilistic programs. Inspired by the uses of time
  credits for reasoning about the running time of deterministic programs, we
  introduce a novel notion of probabilistic cost credit. Probabilistic cost
  credits are a separation logic resource that can be used to pay for the
  cost of operations in programs, and that can be distributed across all
  possible branches of sampling instructions according to their weight, thus
  enabling us to reason about expected cost. The representation of cost
  credits as separation logic resources gives \thetimelogic a great deal of
  flexibility and expressivity. In particular, it permits reasoning about
  amortized expected cost by storing excess credits as potential into data
  structures to pay for future operations. \thetimelogic further supports a
  range of cost models, including running time and entropy usage. We
  showcase the versatility of this approach by applying our techniques to
  prove upper bounds on the expected cost of a variety of probabilistic algorithms
  and data structures, including randomized quicksort, hash tables, and
  meldable heaps.
  All of our results have been mechanized using Coq, Iris, and the
  Coquelicot real analysis library.
\end{abstract}

\maketitle

\section{Introduction}
\label{sec:introduction}

Randomization in programming has many applications, from improving efficiency to ensuring security and privacy.
In many of these applications, it is crucial to reason about the complexity and resource usage of programs.
For example, when analyzing a randomized algorithm, we want to prove that it meets certain performance requirements; and when analyzing security of a cryptographic protocol, standard definitions phrase security in terms of resilience against attacks that run in probabilistic polynomial time.
Since the programs are probabilistic, their running time is typically also a random variable.
One of the most common and important properties to analyze for this random variable is its expected value, known as the \emph{expected running time} of the program.
More generally, we may want to reason about the expected values of other kinds of resource usage, \eg{}, consumed entropy or memory, which leads to a more general notion of \emph{expected cost} of a randomized program.
As in the deterministic setting, sequences of operations on a data structure may alternate between being cheaper and more costly, with the trade-off that a costly operation will allow for many cheaper operations to be performed afterwards.
This leads to the notion of \emph{amortized} cost, which also has its probabilistic variant, known as \emph{amortized expected cost}.

This work presents \thetimelogic{}, a higher-order separation logic for reasoning about (amortized) expected cost of probabilistic programs.
Reasoning about resource usage in separation logic is natural as it is a \emph{logic of resources}.
As pointed out by~\citet{DBLP:journals/corr/abs-1104-1998}, running time can be represented as a separation logic resource through the notion of a \emph{time credit} assertion which can be used to (logically) ``pay'' for steps of execution.
In an affine logic, resources cannot be duplicated, so an initial budget of time credits will provide an upper bound on the number of execution steps.
Prior work has shown that this approach scales to sophisticated algorithms and data structures~\citep{tciris, union-find, pottier_thunks_2024}.

In this work, we build a separation logic that uses credits to reason about \emph{expected} cost.
Reasoning about expected cost poses unique challenges in comparison to reasoning about cost of deterministic programs.
For example, probabilistic programs with overall finite expected cost can still have execution traces with infinite cost.
Consider, \eg{}, the program below:
\begin{align*}
	\langkw{rec}~\langv{coinToss}~\_ = \Tick{1};~\If \Flip then () \Else \langv{coinToss}~()
\end{align*}
Here, $\Flip$ is a command that corresponds to a fair coin toss; it evaluates to either $\True$ or $\False$, each with probability $1/2$.
Meanwhile, the $\Tick{1}$ command is used to represent incurring a cost of $1$ unit of some resource.
Then, the total expected cost of executing $\langv{coinToss}\ ()$ (\ie the expected number of $\Tick{1}$ commands executed) is $2$.

Imagine trying to prove this bound by starting with a precondition of $2$ time credits.
In prior separation logics for deterministic cost bounds, the proof rule for $\Tick{1}$ consumes $1$ time credit.
Thus on every iteration of this function we consume $1$ credit, flip a coin, and then either terminate if the result is $\True$ or do the recursive call if it is $\False$, each with probability $1/2$.
This means that with probability $1/4$ we sample $\False$ on two iterations in a row and consume $2$ credits, leaving us unable to execute any further iterations.
In fact for any initial number of credits $k$ we will need more than $k$ credits to run the program with probability $(1/2)^k$, so no initial amount of credits will suffice.
In order to use credits to reason about the expected running time in this example, we need a way to soundly \emph{increase} the number of credits when sampling $\False$ in the coin toss.

To support this we introduce a new \emph{expectation-preserving} credit rule, inspired by prior work on the expected running time transformer~\cite{kaminski_weakest_2016}.
When reasoning about sampling instructions, this rule allows us to distribute credits across all possible outcomes as long as the expected value over all branches is the initial quantity.
In particular, credits can be scaled up for certain branches, as long as this is compensated for in other branches.
For instance in the example above, after executing $\Tick 1$ we will have $1$ credit, so after tossing the coin we can give $0$ credits to the $\langkw{then}$ branch and $2$ credits to the $\langkw{else}$ branch.
In the former branch we terminate immediately, while in the latter we have the $2$ credits needed to make the recursive call.

Reasoning about \emph{amortized} expected costs using credits does not require any additional reasoning principles.
Instead, amortized reasoning can be achieved by exploiting the expressiveness of separation logic and writing specifications for data structures that store a reserve of credits to pay for more expensive operations, should they arise.
This is in analogy to the \emph{potential method}~\cite{tarjan1985, cormen3rd} often employed in pen-and-paper proofs of amortized complexity of deterministic data structures, and it also generalizes to the probabilistic setting using a similar pattern as appears in prior works based on deterministic time credits~\cite{tciris,union-find}.

Because the analysis of randomized algorithms and data structures often involves different notions of cost, \thetimelogic{} is parameterized by a \emph{cost model} that assigns a cost to every command in the language.
Some previous logics for cost analysis have provided indirect support for encoding cost models through primitives like the $\Tick{n}$ command above, which can be inserted to instrument a program to track some notion of cost.
However, this instrumentation approach works less well for certain natural cost measures for randomized programs, such as \emph{entropy consumed}, where the cost of drawing a random integer uniformly from $\{0, \dots, n\}$ is $\log_2(n+1)$.
In contrast, \thetimelogic' parameterized approach allows us to express this kind of complex cost function in a natural way.

We emphasize that \thetimelogic{} is a fully general program logic that retains all the expressiveness of standard higher-order separation logic.
Furthermore, for deterministic cost-free programs we can reuse existing specifications and proofs without any changes.
By mixing standard separation logic reasoning with reasoning about credits, we achieve greater expressiveness than single-purpose approaches to reasoning about expected cost.

This expressiveness is demonstrated through several case studies, including, a hashmap, a version of Quicksort parameterized by a comparator function, and an algorithm for merging multiple sorted lists, which makes use of a randomized min-heap.
These examples show \thetimelogic' support for higher-order specifications, complex data structure invariants, and compositional reasoning.
Several of these case studies are beyond the scope of prior work for expected cost analysis because they combine higher-order functions with mutable state.
\thetimelogic{} is the first program logic for reasoning about expected costs that supports this combination of features.

\paragraph{Contributions} To summarize, we provide:
\begin{itemize}[topsep=0pt]
\item The first compositional higher-order separation logic, \thetimelogic{}, for reasoning about expected cost of probabilistic programs written in a probabilistic higher-order programming language with higher-order local references.
  \item A probabilistic counterpart of cost credits, which enable us to achieve value-dependent and expectation-preserving proof rules for compositional reasoning.
  \item User-defined cost models, which enable reasoning about a richer class of costs, such as expected entropy use.
  \item A substantial collection of case studies that demonstrates how one can modularly derive non-trivial (amortized) cost bounds of probabilistic programs.
  \item All of the results in this paper have been mechanized in the \rocq proof assistant, using the Iris separation logic framework \cite{irisho, irismonoid, irisjournal, irisinteractive} and the Coquelicot real analysis library \cite{coquelicot}.
\end{itemize}

\paragraph{Outline}
In \cref{sec:key-ideas} we showcase the main features of \thetimelogic by proving expected cost bounds for the program \langv{coinToss} presented earlier and another illustrative example which requires amortized reasoning.
Later in \cref{sec:preliminaries}, we recall some preliminaries and present the language \thelang as well as our notion of cost model.
We then introduce the \thetimelogic logic in \cref{sec:logic}.
We demonstrate how to use \thetimelogic to establish non-trivial expected cost bounds on a range of examples in \cref{sec:case-studies}.
Subsequently in \cref{sec:model-soundness}, we present the model of \thetimelogic and provide a sketch of the adequacy theorem.
Finally, we discuss related and future work in \cref{sec:related-future-work} and conclude in \cref{sec:conclusion}.

\section{Key Ideas}
\label{sec:key-ideas}
\newcommand{\ctprog}{\langv{coinToss}}
\newcommand{\geoprog}{\langv{geo}}

In this section, we demonstrate the features of \thetimelogic by verifying two illustrative examples.
For the first, we show how our expectation-preserving proof rule works. For the second, we show
how representing cost credits as a resource enables us to reason about \emph{amortized}
expected cost for free, without needing to add any other new reasoning principles to our logic.

\subsection{Repeated Coin Toss}

We prove an expected cost bound for the following variant of the coin-tossing program we provided in the introduction
\begin{align*}
  \langkw{rec}~\ctprog~\_ ~=\If \Flip then () \Else \ctprog~()
\end{align*}
where instead of an explicit $\Tick$ operation, we use a cost model that counts every function call, \ie{}, every function application costs $1$ unit while all other reductions have cost $0$ (it is known that just counting the number of function applications gives the right overall asymptotic complexity for programs written in a language like the one considered here~\cite{union-find, ace}). Despite this change, the expected cost is the same, as the cost-inducing operation is just moved from the $\Tick$ to the function application of $\ctprog$.
Hence, it is not surprising that the cost of the program follows a geometric distribution and that the expected cost is $2=\sum_{k=0}^{\infty}2^{-k}$. We thus use \thetimelogic{} to prove the following bound:
\[\ec{\ctprog~(),\emptyset} \leq 2\]
where $\ec{e, \sigma}$ denotes the expected cost of running the program $e$ with heap $\sigma$, and $\emptyset$ denotes the empty initial heap.
In \cref{sec:logic} we will discuss the \textit{adequacy theorm} for \thetimelogic{}, which states that if we prove a Hoare triple for $\expr$ in \thetimelogic{} and we ``own $x$ credits in the beginning'', then the expected cost of running $\expr$ is less than $x$.
Hence it suffices to prove the following specification:
\begin{align}
  \hoare {\tc {2}}{\ctprog~\TT}{\TRUE} \label{eqn:ct-spec}
\end{align}
where the assertion $\tc x$ in the precondition expresses that we start our proof owning $x$ credits.

We first note that $\ctprog$ is recursive, and so we apply a proof rule for recursive function applications. Under the current cost model function applications cost $1$ unit, so this rule has the following form:
\begin{mathpar}
    \infrule{ht-rec}
	  {{\All \valB . \hoare{\prop \sep \tc{1}}{(\Rec f x = e)\ \valB}{\propB}}\vdash \hoare{\prop}{\subst{\subst{e}{x}{\val}}{f}{(\Rec f x = e)}}{\propB} }
	  {\vdash \hoare{\prop\sep \tc{1}}{(\Rec f x = e)\ \val}{\propB}}
\end{mathpar}
This rule is relatively standard, except for the additional $\tc{1}$ credits in the precondition, which we need in order to ``pay'' for the function application.
In particular, it says that we may assume the desired specification for any recursive calls (the assumption to the left of the $\vdash$) while reasoning about the body of the function.

Since we are considering a cost model where all reductions other than function application incur no costs, we have ``standard'' proof rules for non-function application steps, \eg, the following rule.
\begin{mathpar}
    \infrule{ht-if}
    {\vdash \hoare{\prop\sep b=\True}{\expr_1}{\propB} \\ \vdash \hoare{\prop\sep b=\False}{\expr_2}{\propB} }
    {\vdash \hoare{\prop}{\If b then \expr_1 \Else \expr_2}{\propB}}
\end{mathpar}

Coming back to our example, note that instead of exactly $\tc{1}$ credits for the precondition of the \rref{ht-rec} rule, we have $\tc{2}$.
Hence we need to split our credits up to pay for this function application step via the following rule for credits $\tc{x_1} \sep \tc{x_2} \dashv \vdash \tc{x_1 + x_2}$.
Thus we split $\tc{2}$ into $\tc{1}\sep\tc{1}$, where the first $\tc{1}$ is used to pay for the function application and the second is passed on to reason about the remainder of the program (the body of the function call).
After splitting the credits and applying \rref{ht-rec} to \cref{eqn:ct-spec}, we are left to prove
\begin{align*}
	 \hoare {\tc{1} \sep \tc{1} }{\ctprog~\TT}{\TRUE} \vdash\hoare {\tc {1}}{\If \Flip then () \Else \ctprog~\TT}{\TRUE}
\end{align*}

Next, we reason about the coin toss (the $\Flip$).
We need to consider the expected cost for the cases where the toss returns $\False$ and where it returns $\True$.
We observe that the two cases have different expected cost; therefore, we need to split our credits in an ``expectation-preserving'' manner across the two branches, as alluded to in the introduction.
To do so, we use the \emph{expectation rule} for $\Flip$:\footnote{We provide a generalized expectation rule for more general uniform sampling in \cref{sec:proof-rules}.}
\begin{mathpar}
  \infrule{ht-flip-exp}{
	  \tfrac{1}{2}\cdot \Rt_2(\True) + \tfrac{1}{2}\cdot \Rt_2(\False) \leq \rt_1
  }{
    \hoare{\tc{\rt_1}} {\Flip} {b \ldotp \tc{\Rt_2(b)}}
  }
\end{mathpar}
This rule states that for $\Rt_{2} : \bool \to \nnreal$ we may split $\tc{\rt_1}$ into $\tc{\Rt_2(\False)}$ and $\tc{\Rt_2(\True)}$ for the two branches of $\Flip$, if their weighted sum is less than or equal to $\rt_1$.
This encodes the intuition that we distribute credits in a expectation-preserving way, where the expected value over all branches is bounded above by the initial quantity.
Applying this rule in our example, from the $\tc{1}$ in the precondition we can pass $\tc{0}$ to the $\langkw{then}$ branch and $\tc{2}$ to the $\langkw{else}$ branch.
That is, we pick $\Rt_2(b) \eqdef{} \textit{if}~b~\textit{then}~0~\textit{else}~2$ which is a valid choice since $\frac{0}{2} + \frac{2}{2} = 1 \le 1$.
By \rref{ht-flip-exp} and an application of the standard \textsc{ht-bind} rule, it then suffices to show:
\begin{align*}
  \hoare {\tc 1 \sep \tc 1}{\ctprog~\TT}{\TRUE} &\vdash
  \hoare{
    \tc{\Rt_2(b)}
  }{\If b then () \Else \ctprog\ ()}
        {\TRUE}
\end{align*}
From here, we can apply the \rref{ht-if} rule, which performs a case split on $b$ to verify the rest of the two branches depending on the return value of $\Flip$, \ie,
\begin{align*}
  \hoare {\tc 1 \sep \tc 1}{\ctprog\ ()}{\TRUE} &\vdash
  \hoare{
    \tc{\Rt_2(b)} \sep b=\True
  }{()}
        {\TRUE}\\
        \hoare {\tc 1 \sep \tc 1}{\ctprog\ ()}{\TRUE} & \vdash
        \hoare{
          \tc{\Rt_2(b)} \sep b=\False
        }{\ctprog\ ()}
              {\TRUE}
\end{align*}
By symbolic simplification of the preconditions above, it thus suffices to show
\begin{align*}
  \hoare {\tc 1 \sep \tc 1}{\ctprog\ ()}{\TRUE} &\vdash
  \hoare{
    \tc{0}
  }{()}
        {\TRUE}\\
        \hoare {\tc 1 \sep \tc 1}{\ctprog\ ()}{\TRUE} & \vdash
        \hoare{
          \tc{2}
        }{\ctprog\ ()}
              {\TRUE}
\end{align*}
The first program immediately returns $()$, which trivially satisfies the postconsition $\TRUE$, without incurring any additional cost, finishing the case.
For the second, we apply the assumption after splitting the $\tc 2$ credits as $\tc 1 \sep \tc 1$.

If one chooses to count only the number of calls to $\Flip$, one can prove the same specification in a similar manner.
Indeed, the proof rules and soundness of \thetimelogic are agnostic to the definition of the cost model,
and users can define their own cost model, as we shall see in \cref{sec:cost-models}.

To summarize, we demonstrated how one uses \thetimelogic to verify the expected cost of a simple coin-tossing program. In this one example, we highlighted how one can
\begin{enumerate}
\item Use credits, a separation logic resource that enjoys all the usual separation logic rules (such as the frame rule), to keep track of the cost incurred as we step through the program.
\item Use expectation-preserving composition to reason about the expected cost across multiple branches after sampling operations.
\end{enumerate}

\subsection{Repeated Amortized Operations}
\label{sec:repeated-amortized-operations}
\newcommand{\amortizedop}{\mathsf{op}}
\newcommand{\repeatN}{\mathsf{repeat}}
Next, we consider an example of amortized expected time reasoning introduced by
\citet{DBLP:journals/pacmpl/BatzKKMV23} to motivate their amortized ERT calculus.
We will see that just by combining credits with standard separation logic reasoning rules, we can handle the amortized reasoning in this example.

Let $\expr_{1} \oplus \expr_{2}$ be notation for $\If \Flip then \expr_{1} \Else \expr_{2}$ and consider the program
\begin{align*}
  \amortizedop \eqdef{} (\Tick(\deref \loc) \oplus \loc \gets 0); \loc \gets (\deref \loc) + 1
\end{align*}
Here the operator $\Tick(\deref \loc)$ indicates a cost of $\deref\loc$, where $\loc$ is some memory location and $\deref$ is
the dereferencing operator.
If location $\loc$ contains a natural number $n$ then with probability $\frac{1}{2}$, $\amortizedop$ either has cost $n$ or resets $\loc$ to $0$.
In both cases, $\loc$ is subsequently incremented.
As pointed out by \citeauthor{DBLP:journals/pacmpl/BatzKKMV23}, it is easy to show that if $\loc$ contains $n$, the expected cost of the operation is $\frac{n}{2}$, \ie{},
\begin{align*}
  \hoare
  {\tc{\tfrac{n}{2}} \sep \loc \mapsto n}
  {\amortizedop}
  {\Exists m . \loc \mapsto m}
\end{align*}
where $\loc \mapsto n$ is the standard \emph{points-to} connective from separation logic stating that $\loc$ contains the value $n$.
However, computing the cost of a \emph{sequence} of operations in general is much less obvious.
For example, one can show that the expected cost of $(\amortizedop; \amortizedop)$ is $\frac{3n}{4} + \frac{1}{4}$ and for $(\amortizedop; \amortizedop; \amortizedop)$ it is $\frac{7n}{8} + \frac{5}{8}$ which does not lead to an obvious pattern.
However, repeating the operation $m$ times has \emph{amortized} cost $n + m$, which \citeauthor{DBLP:journals/pacmpl/BatzKKMV23} establish using a dedicated amortized expected running time calculus.
In \thetimelogic, such amortized expected reasoning emerges from using credits with standard separation logic reasoning, just as prior work has shown that deterministic credits enable deterministic amortized bounds~\citep{DBLP:journals/corr/abs-1104-1998, union-find}.

To show the amortized specification, we first establish that $\amortizedop$ has amortized cost $1$, \ie{}, we show
\begin{align*}
  \hoare
  {\tc{1} \sep \pi}
  {\op}
  { \_ \ldotp \pi}
\end{align*}
where $\pi = \Exists n . \loc \mapsto n \sep \tc{n}$.
Intuitively, $\pi$ contains the \emph{potential} needed to evaluate the operation which changes according to the contents of $\loc$.
The precondition of the specification thus gives us a total of $n + 1$ credits, where $n$ is the content of $\loc$.
The key part of the proof is applying the \rref{ht-flip-exp} rule using $\Rt_2(b) \eqdef{} \textit{if}~b~\textit{then}~2n + 1~\textit{else}~1$.
If $\Flip$ evaluates to $\False$, we reset $\loc$ to 0 and thus no longer need the $n$ credits stored in $\pi$ for that branch.
If $\Flip$ evaluates to $\True$, we have to both spend $n$ credits and also maintain $\pi$ with $n$ credits, so we need $2n$ credits in total.
As $\frac{2n+1}{2} + \frac{1}{2} = n + 1$, the expectation rule allows us, intuitively, to bring the $n$ credits from the $\langkw{else}$ branch to the $\langkw{then}$ branch, as the credits needed in the two branches average out.
In both cases, we will have the $\tc 1$ supplied by the precondition left over, which we need to store back as potential, since the content of $\loc$ is incremented and this re-establishes $\pi$ as a postcondition.

It is now straightforward to show an amortized specification of $m$ repetitions of $\amortizedop$.
Let $\repeatN~m~f$ be a program that invokes the $f$ procedure $m$ times.
By applying a general higher-order specification of $\repeatN$ and the specification of $\amortizedop$ we show
\begin{align*}
  \hoare
  {\tc{n + m} \sep \loc \mapsto n}
  {\repeatN~m~(\Lam \_ . \amortizedop)}
  {\TRUE}
\end{align*}
as a simple corollary and thus an amortized expected cost of $n + m$.

To summarize, this example illustrates how the resource-based representation
plus the expressivity of the ambient higher-order separation logic enables us to reason about
amortized expected cost without the need for any other specialized rules. In general, by ``depositing''
excess credits in a data structure's potential assertion (like the $\pi$ above), and then ``withdrawing''
those stored credits to pay down expensive operations, we can capture standard patterns of amortized reasoning.
The \rref{ht-flip-exp} rule further allows us to average stored credits across branches to do amortized expected reasoning.

\section{Preliminaries}
\label{sec:preliminaries}
In this section, we first recall some useful definitions from probability theory. We then present the syntax of \thelang, the high-order probabilistic ML-like language our programs are written in, as well as  its operational semantics. Lastly, we present the notion of cost models, the device that allows us to define a more general class of costs depending on the application, and we define the expected cost of a program according to a cost model. 

\subsection{Probabilities}
\label{sec:probabilities}

A \defemph{discrete subdistribution} (henceforth simply \defemph{distribution}) on a countable set $A$ is a function $\distr : A \ra [0,1]$ such that $\sum_{a \in A}\distr(a) \leq 1$. The collection of distributions on $A$ is denoted by $\Distr A$.

A (non-negative, real) \defemph{random variable} with respect to a distribution $\distr : \Distr A$ is a function $\rv : A \ra \nnreal$, where $\nnreal$ denotes the non-negative real numbers.
The \defemph{expectation} or \defemph{expected value} of a random variable $\rv$ \wrt $\distr$ is defined as $\expect[\distr] \rv \eqdef{} \sum_{a \in A} \distr(a) \cdot \rv(a)$
if the sum converges, in which case we say that $\rv$ has finite expectation.

The collection of distributions can be equipped with a monad structure as usual:
\begin{lemma}[Discrete Distribution Monad]
  Let $\mu \in \Distr{A}$, $a \in A$, and $f : A \to \Distr{B}$.
  Then\vspace*{-1mm}
  \begin{align*}
    \mbind(f,\mu)(b) \eqdef{} \sum_{a \in A} \mu(a) \cdot f(a)(b)
    &&
    \mret(a)(a') \eqdef{}
            \begin{cases}
              1 & \text{if } a = a' \\
              0 & \text{otherwise}
            \end{cases}
  \end{align*}
  gives a monadic structure to $\DDistr$.
  We write $\mu \mbindi f$ for $\mbind(f, \mu)$.
\end{lemma}

\subsection{Language Definition and Operational Semantics}
\label{sec:prelim-opsem}

The syntax of $\thelang{}$, the language we consider in this paper, is defined by the grammar below.
\begin{align*}
  \val, \valB \in \Val \bnfdef{}
                             & z \in \integer \ALT
  b \in \bool \ALT
  \TT \ALT
  \loc \in \Loc \ALT
  \Rec f x = \expr \ALT
  (\val,\valB) \ALT
  \Inl \val  \ALT
  \Inr \val
  \\
  \expr \in \Expr \bnfdef{}  &
  \val \ALT
  \lvar \ALT
  \Rec f x = \expr \ALT
  \expr_1~\expr_2 \ALT
  \expr_1 + \expr_2 \ALT
  \expr_1 - \expr_2 \ALT
  \ldots \ALT
  \If \expr then \expr_1 \Else \expr_2 \ALT \\
  &(\expr_1,\expr_2) \ALT
  \Fst \expr \ALT
  \Snd \expr \ALT
  \Inl \expr \ALT
  \Inr \expr \ALT
  \Match \expr with \Inl \val~ => \expr_1 | \Inr \valB => \expr_2 end \ALT  \\
  & \AllocN~\expr_1~\expr_2 \ALT
  \deref \expr \ALT
  \expr_1 \gets \expr_2 \ALT
  \expr_1 [\expr_2] \ALT
  \Rand \expr \ALT
  \Tick \expr
  \\
  \lctx \in \Ectx \bnfdef{}  &
  -
  \ALT \expr\,\lctx
  \ALT \lctx\,\val
  \ALT \AllocN~\expr~\lctx
  \ALT \AllocN~\lctx~\val
  \ALT \deref \lctx
  \ALT \expr \gets \lctx
  \ALT \lctx \gets \val
  \ALT \Rand \lctx
  \ALT \Tick \lctx
  \ALT \ldots
  \\
  \state \in \State \eqdef{} & \Loc \fpfn \Val
                               \qquad \qquad
                               \cfg \in \Conf \eqdef{} \Expr \times \State
\end{align*}
The term language is mostly standard:
$\AllocN \expr_1~\expr_2$ allocates a new array of length $\expr_1$ with each cell containing the value returned by $\expr_2$, $\deref \expr$ dereferences the location $\expr$ evaluates to, and $\expr_{1} \gets \expr_{2}$ assigns the result of evaluating $\expr_{2}$ to the location that $\expr_{1}$ evaluates to.
We also write $\Aoff \loc [i]$ for offsetting location $\loc$ by $i$.
Reading from and writing $\val$ to an array $a$ at offset $i$ is thus written as $\deref \Aoff a[i] $ and $\Aoff a[i] \gets \val $ respectively.
We often refer to a recursive function value $\Rec f x = \expr$ by its name $f$.
We introduce syntactic sugar for lambda abstractions $\Lam \var . \expr$ defined as $\Rec {\_} \var = \expr$,
let-bindings $\Let \var = \expr_{1} in \expr_{2}$ defined as $(\Lam \var . \expr_{2})~\expr_{1}$,  sequencing $\expr_{1} ; \expr_{2}$ defined as $\Let \_ = \expr_{1} in \expr_{2}$, and references $\Alloc \expr$ defined as $\AllocN~1~\expr$. 
The Boolean operation $\Flip$ is syntactic sugar for $(\Rand 1 == 1)$.
The $\Tick$ instruction has no operational meaning ($\Tick n$ simply reduces to~$\TT$) but can be used by cost models to explicitly specify the cost of a computation in the program, see, \eg{}, $\costtick$ in \cref{sec:cost-models}.

To define full program execution, we define $\stepdistr(\cfg) \in \Distr{\Conf}$, the distribution induced by the single step reduction of configuration $\cfg \in \Conf$. The semantics is standard,
we first define head reductions and then lift it to reduction in an evaluation context $K$.
All non-probabilistic constructs reduce deterministically as usual, \eg, $\stepdistr(\If \True then \expr_1 \Else \expr_2, \sigma) = \mret(\expr_{1}, \sigma)$ and $\stepdistr(\Tick z, \sigma) = \mret(\TT, \sigma)$. The probabilistic choice $\Rand \tapebound$ reduces uniformly at random, \ie,
\begin{align*}
	& \stepdistr(\Rand \tapebound, \sigma)(n, \sigma) \eqdef{}
  \begin{cases}
    \frac{1}{\tapebound + 1} & \text{for } n \in \{ 0, 1, \ldots, N \}, \\
    0                        & \text{otherwise}.
  \end{cases}
\end{align*}

With the single step reduction $\stepdistr(-,-)$ defined, we now define a stratified execution probability $\exec_{n}\colon \Conf \to \Distr{\Val}$ by induction on $n$:
\begin{align*}
 \exec_{n}(\expr, \state) \eqdef{}
 \begin{cases}
   \Lam \val . 0                                           & \text{if}~\expr \not\in\Val~\text{and}~n = 0, \\
   \mret(\expr)                                         & \text{if}~\expr \in \Val, \\
   \stepdistr(\expr, \state) \mbindi \exec_{(n - 1)} & \text{otherwise.}
 \end{cases}
\end{align*}
The probability that a full execution, starting from configuration $\cfg$, reaches a value $\val$ is taken as the limit of its stratified approximations, which exists by monotonicity and boundedness:
\begin{align*}
 \exec(\cfg)(\val) \eqdef{} \lim_{n \to \infty} \exec_{n}(\cfg)(\val)
\end{align*}
We simply write $\exec(e)$ as notation for $\exec (e,\sigma)$ if $\exec(e,\sigma)$ is the same for all states $\sigma$.
The termination probability of an execution from a configuration $\cfg$ is denoted by $\execTerm{(\cfg)} \eqdef{} \Sigma_{\val\in\Val}\exec(\cfg)(v)$.

\subsection{Cost Models and Expected Cost}
\label{sec:cost-models}

A \defemph{cost model} is a non-negative real-valued function $\cost : \Expr \rightarrow \nnreal$ that is invariant under evaluation contexts in the sense that if $\expr \not\in \Val$ then $\cost (\fillctx \lctx[\expr]) = \cost(\expr)$ for all $\lctx$ and $\expr$.
We use cost models to associate costs (non-negative real values) to one step of evaluating an expression, \ie, $\cost(e)$ represents the cost incurred by the next transition of $e$.
At its simplest, a cost model can associate a constant cost to all expressions, and hence to all reduction steps.
For instance, the cost model $(\Lam \_ . 1)$ is obviously invariant under evaluation contexts.

For a given cost model $\cost$, the \defemph{expected cost} of executing $\expr$ for $n$ steps from starting state $\state$ is defined as follows.
\begin{align}
  \label{def:ertstep}
  \ec[n][\cost]{\expr, \state}
	&\eqdef{}
    \begin{cases}
      0 & \text{if } n=0 \text{ or } \expr \in\, \Val, \\
      \cost(\expr) + \expect[\stepdistr(\expr,\state)]{{\EC_m^{\cost}}}
        & \text{if $n = m+1$.}
    \end{cases}
\end{align}
In this definition, we consider $\EC_m^\cost$ as a random variable \wrt the distribution induced by stepping the current configuration.
Concretely, $\expect[\stepdistr(\expr,\state)]{{\EC_m^{\cost}}} = \sum_{\cfg \in \Conf} \stepdistr(\expr,\state)(\cfg) \cdot \ec[m][\cost] \cfg$.
Note that our language semantics ensures that $\stepdistr(\expr,\state)(\cfg)$ is non-zero at finitely many configurations,
so this expected value is always finite.

The \defemph{expected cost} for full program execution is defined as
$\EC^{\cost} \eqdef{} \sup_{n \in \omega} \EC_n^{\cost}$, where we take
$\EC^{\cost} \eqdef{} \infty$ if the sequence is not bounded.

\paragraph{Examples of cost models}
Since cost models have to respect evaluation contexts, they are usually defined in term of the current head redex of an expression. Given a (reducible) expression $\expr$, there exists a unique maximal evaluation context $\lctx$ and expression $\expr'$ such that we can decompose $\expr$ as $\fillctx \lctx[\expr']$ with $\expr'$ reducible using a head reduction step.
Let $\decomp : \Expr \rightarrow \Expr$ denote the function that computes $\expr'$ from $\expr$.
One can show that if a cost model $\cost$ is defined in terms of $\decomp$, then $\cost$ automatically satisfies the requirement of context-invariance.

For example, we can define the following cost functions.
\begin{align*}
  \costall &\eqdef{} \Lam \_ . 1
  \\
  \costapp &\eqdef{} \Lam \expr . 1 && \text{if } \decomp(\expr) = \expr_1\ \expr_2 &&\text{ for some } \expr_1, \expr_2 \text{,\; and } &0\quad \text{otherwise.}
  \\
  \costrand &\eqdef{} \Lam \expr . \log_2 (\tapebound+1) && \text{if } \decomp(\expr) = \Rand \tapebound &&\text{ for some } \tapebound \text{,\; and } &0\quad \text{otherwise.}
  \\
  \costtick &\eqdef{} \Lam \expr . |z| && \text{if } \decomp(\expr) = \Tick z &&\text{ for some } z \in \integer \text{,\; and } &0\quad \text{otherwise.}
\end{align*}

As previously mentioned, $\costall$ associates a constant cost to all reduction steps.
The cost model $\costapp$ was used in the example in \cref{sec:key-ideas}, and counts the number of function applications. This is known to behave asymptotically
as the total number of reduction steps, but it has the advantage that it allows to abstract away some concrete implementation details.
$\costrand$ counts the entropy cost incurred by sampling commands. Recall that sampling uniformly out of a set of $N$ elements
has an entropy cost of $\log_2 N$.
Generating high-quality random bits can be expensive, particularly if they must be cryptographically secure, so entropy usage of a computation is an important cost consideration~\cite{DBLP:journals/corr/abs-1304-1916}.
Finally, $\costtick$ counts the cost of all $\Tick$ instructions. This is in line with previous works, and allows
for cost analysis through code annotations.

\section{The Expected Cost Logic}
\label{sec:logic}
The \thetimelogic logic for expected cost is built on top of the Iris base logic \cite{irisjournal}, a state-of-the-art higher-order separation logic.
On top of the base logic, we add a predicate that denotes ownership of credits and a notion of Hoare triple that can be used to write specifications of programs.
Hoare triples can then be established using a set of program logic rules and mapped to a concrete bound on expected cost using an adequacy theorem.

In this section we discuss each of these aspects in more detail but save the proof of soundness for \cref{sec:model-soundness}.
Below, we parameterize the development in terms of an arbitrary cost function called $\cost$ that satisfies the conditions discussed in \cref{sec:cost-models}.

\subsection{Base Logic of Propositions}
\label{sec:base-logic}

The Iris base logic is an expressive higher-order separation logic framework with support for advanced features such as higher-order ghost state \cite{irisho}, impredicative invariants \cite{icap}, and guarded recursion, all of which are paramount for reasoning about higher-order programs written in realistic languages.
While important, we only discuss these more specialized connectives as necessary to focus on the novelties of this work.
A selection of \thetimelogic propositions are shown below.
\begin{align*}
  \prop,\propB \in \iProp \bnfdef{}
  & \TRUE \ALT \FALSE \ALT \prop \land \propB \ALT \prop \lor \propB \ALT \prop \Ra \propB \ALT
    \All \var . \prop \ALT \Exists \var . \prop \ALT\\
  &\prop \sep \propB \ALT \prop \wand \propB \ALT 
    \progheap{\loc}{\val} \ALT
    \tc{\rt} \ALT
    \hoare{\prop}{\expr}{\val\ldotp\propB} \ALT
    \ldots
\end{align*}
With the exceptions of the credit assertion $\tc{\rt}$ and Hoare triples $\hoare{\prop}{e}{v.\propB}$, all propositions have the usual meaning as in (non-probabilistic) separation logic.
The $\tc{\rt}$ assertion denotes ownership of $\rt \in \nnreal$ credits.
This is a separation logic resource subject to a set of rules, the most important of which is the credit splitting rule $\tc{\rt_1 + \rt_2} \dashv \vdash \tc{\rt_1} \sep \tc{\rt_2}$.
That is, owning $\rt_1 + \rt_2$ credits is equivalent to owning both $\rt_1$ and $\rt_2$ credits.
Splitting credits allows us, \eg{}, to pass a subprogram the exact amount of credits it needs while keeping the remaining credits in reserve. 
The logic is affine, which means that we may ``throw away'' resources and credits in particular, but they cannot be duplicated or generated from nothing.
This ensures that the initial amount of credits assumed by a program provides a sound upper bound on its cost.

A Hoare triple $\hoare{\prop}{\expr}{\val\ldotp\propB}$ is a proposition that states a specification for a program $\expr$.
Here $\prop$ denotes the precondition and $\propB$ a postcondition with a free variable $v$, which morally captures the return value of $\expr$.
The \thetimelogic{} logic is \emph{impredicative}, \ie{}, both Hoare triples and their pre- and postconditions are first-class propositions of the logic, and in particular they may contain Hoare triples as well.
This is important for giving general and compositional specifications to higher-order programs as showcased in \cref{sec:case-studies}.
In order to discuss the examples in \cref{sec:case-studies}, and for any user of \thetimelogic{}, it is only important to understand what Hoare triples imply (through the adequacy theorem) and how they are proven (using the program logic rules) rather than how they are defined in \cref{sec:model-soundness}.

\subsection{Program Logic}
\label{sec:proof-rules}

Soundness of the program logic is captured by the following adequacy theorem.
\begin{theorem}[Adequacy]\label{thm:adequacy-hoare}
  Let $\rt$ be a non-negative real number and let $\varphi$ be a predicate on values. If~$\vdash \hoare {\tc \rt} \expr \varphi$ then for any state $\state$,
  \begin{enumerate}
  \item \label{thm:wp-adequacy-cost} $\ec{\expr,\state} \leq \rt$, and
  \item \label{thm:wp-adequacy-correctness} $\All \val \in \Val\, . \exec(\expr,\state)(\val) > 0 \implies \varphi(\val)$.
  \end{enumerate}
\end{theorem}
The theorem says that by proving a Hoare triple specification of $\expr$ that assumes initial ownership of $\rt$ credits, then the expected cost of executing $\expr$ is bounded above by $\rt$.
Moreover, any value in the output distribution satisfies $\varphi$.
Recall that we assume an ambient $\cost$ function which the concrete definition of $\ec{\cdot, \cdot}$ depends on.

\Eclog Hoare triples satisfy a range of structural and computational rules.
Notably, both the structural rules and the computational rules look almost identical to the corresponding rules found in separation logics for reasoning about non-probabilistic programs, with the addition of cost tracking through credits.
We emphasize that this is a strength of our approach and proofs of non-probabilistic programs, \eg{}, data structures, can readily be reused if the cost model permits.

A selection of structural and computational rules are shown below.
\begin{mathpar}
  \infrule{ht-bind}
  {\vdash \hoare{\prop}{\expr}{\val . \propB} \\ \vdash \All \val . \hoare{\propB}{\fillctx\lctx[\val]}{\propC}}
  {\vdash \hoare{\prop}{\fillctx\lctx[\expr]}{\propC}}
  \and
  \infrule{ht-frame}
  {\vdash \hoare{\prop}{\expr}{\propB}}
  {\vdash \hoare{\prop \sep \propC}{\expr}{\propB \sep \propC }}
  \and
  \infrule{ht-load}
  { }
  {\proves \hoare{\progheap{\loc}{\val} \sep \tc{\cost(\deref\loc)}}{\deref\loc}{\Ret \valB . \valB = \val \sep \progheap{\loc}{\val}}}
  \and
  \infrule{ht-rand}
  { }
  {\proves \hoare{\tc{\cost(\Rand N)}}{\Rand N}{\Ret n . 0 \leq n \leq N}}
  \and
  \infrule{ht-rec}
  { \expr_f = (\Rec f x = \expr)
    \\
    \All \valB . \hoare{\prop \sep \tc{\cost(\expr_f~\valB)}}{\expr_f~\valB}{\propB} \vdash
    \hoare{\prop}{\subst{\subst{e}{x}{\val}}{f}{\expr_f}}{\propB} }
  {\vdash \hoare{\prop \sep \tc{\cost(\expr_f~\val)}}{\expr_f~\val}{\propB}}
\end{mathpar}
Notice in particular that \ruleref{ht-bind} is the ``usual'' bind rule.
This fact relies on the requirement that cost models are invariant under evaluation contexts, \ie{}, the cost of evaluating $\expr$ in context $\lctx$ is the same as the cost of evaluating $\expr$ in isolation.
The frame rule \ruleref{ht-frame} is the traditional separation-logic frame rule.
Since credits are assertions, they can be framed using this rule just like any other separation logic resource.

The computational rule \ruleref{ht-load} is used to reason about read operations.
It requires ownership of the location $\loc$ as denoted by the points-to connective $\progheap{\loc}{\val}$.
In addition, we are required to own $\cost(\deref\loc)$ credits to pay for the (possible) cost of the read operation.
Similar computational rules hold for store and allocation operations.
The rule \rref{ht-rec} can be used to reason about recursive functions.
The rule requires ownership of enough credits to pay for the function invocation and provides a Hoare triple specification for reasoning about recursive calls of $\expr_f$.
While the rule allows the cost to depend on the actual function and its arguments, we only consider cost models that assign a uniform cost to all function applications.

The sampling operation $\Rand N$ satisfies a computational rule \ruleref{ht-rand} similar to \ruleref{ht-load} where $\cost(\Rand N)$ credits are required in the precondition.
However, a key novelty of this work is that it also satisfies the rule \ruleref{ht-rand-exp} shown below which allows for \emph{expectation-preserving composition}.
The rule distributes credits among all possible outcomes of a sampling in an expectation-preserving manner.
Specifically, suppose we are given $\rt_1$ credits in the precondition of a $\Rand N$ operation which samples uniformly from the set $\{0,\dots, N\}$.
Given that $n$ is the outcome of the sampling, the rule says that we can supply $\Rt_2(n)$ credits for the remainder of the program, as long as the cost of $\Rand N$ and the \emph{weighted} sum of $\Rt_2:\{0,\dots, N\} \rightarrow \nnreal$ is at most $\rt_1$.
\begin{mathpar}
  \infrule{ht-rand-exp}{
    \cost(\Rand N) + \textstyle\sum_{n = 0}^N \frac{\Rt_2(n)}{N+1} \leq \rt_1
  }{
    \hoare{\tc{\rt_1}} {\Rand N} {\Ret n . \tc{\Rt_2(n)} \sep {0 \leq n \leq N}}
  }
\end{mathpar}

\subsection{Tactic Support}

One might worry that the $\tc{\cost(e)}$ preconditions in all of the rules add significant overhead to carrying out proofs in \thetimelogic{}, even in a cost model where most of these costs are $0$.
To address this, the Coq formalization of \thetimelogic{} extends the Iris Proof Mode~\cite{mosel} which provides tactic support for interactively reasoning about judgments in the Iris separation logic.
For example, when verifying a program and applying \ruleref{ht-load}, the tactics automatically decompose the expression in the goal into its evaluation context and head redex, apply \ruleref{ht-bind}, and locate the corresponding points-to connective in the resource context before applying \ruleref{ht-load}.
We adapt these tactics to automatically determine the cost of the redex and locate sufficient credits in the resource context.
For the cost models we have considered, managing credits is no more of a nuisance than managing points-to connectives, even for models like $\costall$ where credits must be provided for all execution steps.

\section{Case Studies}
\label{sec:case-studies}

In this section, we present a collection of case studies where we apply \thetimelogic to reason about expected cost.
The examples come from a spectrum of applications, each presenting various challenges, \eg{}, amortized reasoning, local state, and compositionality.
We demonstrate the usability of \thetimelogic by proving concise, non-trivial, and modular specifications for each of these examples.

\subsection{Coupon Collector}
\label{sec:coupon-collector}
\newcommand{\hsum}{H}
\newcommand{\arrvar}{\langv{arr}}
\newcommand{\cntvar}{\langv{cnt}}
\newcommand{\drawvar}{\langv{k}}
\newcommand{\ccprog}{\langv{collector}}
\newcommand{\cchprog}{\langv{repeatDraw}}
\newcommand{\llen}{\langv{length}}
\newcommand{\ssize}{\langv{size}}
\newcommand{\tset}{s}
\newcommand{\arrl}{l}
\newcommand{\lsmatch}{\langv{lsmatch}}

The coupon collector's problem is a famous puzzle in probability theory.
The problem goes as follows: given $n>0$ distinct coupons in a bag, what is the expected number of random draws with replacement you need to do to draw every coupon at least once?
A standard pen-and-paper analysis starts by observing that the total number of draws $D$ can be rewritten as the sum $D=\sum_{i=1}^n D_i$ where $D_i$ is the random variable representing the number of draws to draw the $i$th coupon after having drawn $i-1$ distinct coupons previously.
By linearity of expectation, we have $\expect{D}=\expect{\sum_{i=1}^n D_i}=\sum_{i=1}^n \expect{D_i}$.
Then, each $D_i$ follows a geometric distribution with expectation $\frac{n}{n-(i-1)}$ and hence $\expect{D}=\sum_{i=1}^n \frac{n}{n-(i-1)}=n\cdot(\sum_{i=1}^n \frac{1}{i})=n\cdot \hsum(n)$ where $\hsum (x)$ represents the harmonic sum $\sum_{i=1}^x \frac{1}{i}$.
As we shall see, the frame rule allows us to carry out a proof in \thetimelogic{} that is quite similar in structure to this standard analysis.

We implement the coupon collection process as the \thelang program shown in \cref{fig:coupon}.
The implementation initially allocates an array $\arrvar$ of length $n$ where each cell is initialized to $\False$; the array tracks the coupons that have been collected so far.
The program repeatedly samples a coupon $\drawvar$ by executing $\Rand~(n-1)$.
If $\arrvar[\drawvar]$ is $\True$, the coupon has already been drawn, so the process repeats.
Otherwise, we mark $\arrvar$ and repeat the process with $\cntvar$ decremented by $1$.
The process stops when $\cntvar$ reaches 0 which signifies that $n$ distinct coupons have been drawn.
\begin{figure*}[t]
  \centering
  \begin{minipage}[t]{.45\linewidth}
    \begin{align*}
      &\Rec \cchprog \arrvar~\cntvar = \\
      &\quad\If \cntvar = 0 then ()\\
      &\quad\Else
        \begin{aligned}[t]
          &\Let \drawvar = \Rand~(n-1) in \\
          &\If \deref \arrvar[\drawvar] then \cchprog\ \arrvar\ \cntvar\\
          &\Else
            {\begin{aligned}[t]
               &\arrvar[\drawvar] \gets \True; \\
               &\cchprog\ \arrvar~(\cntvar - 1)
             \end{aligned}}
        \end{aligned}
    \end{align*}
  \end{minipage}
  \begin{minipage}[t]{.45\linewidth}
    \begin{align*}
      \ccprog \eqdef{} &\Lam \_. \\
                       &\Let \arrvar = \AllocN\ n\ \False in\\
                       &\cchprog\ \arrvar\ n
    \end{align*}
  \end{minipage}
  \caption{An implementation of the coupon collector.}
  \label{fig:coupon}
\end{figure*}

Following the analysis shown above, we expect to prove the following specification using a cost model that assigns cost 1 to $\Rand N$ and $0$ to everything else.
\begin{align*}
  \hoare {\tc {n\cdot \hsum(n)}}{\ccprog\ ()}{\TRUE}
\end{align*}
We start by proving a specification for the $\cchprog$ loop.
\begin{align*}
  \hoare{
  \begin{array}{l}
    0 < \cntvar\le n \sep 
    \llen\ \arrl = n \sep
    \ssize\ \tset = n-\cntvar \sep{} \\ 
    \lsmatch\ \arrl\ \tset \sep
    \tc{n\cdot \hsum(\cntvar)} \sep
    \arrvar \mapsto^\ast \arrl
  \end{array}}
  {\cchprog~\arrvar~\cntvar}
  {\TRUE}
\end{align*}
Here $\arrvar \mapsto^\ast \arrl$ denotes ownership of an array starting at location $\arrvar$ and the list of its contents $l$.
The predicate $\lsmatch$ asserts that that $i$th index of the list $\arrl$ is $\True$ if and only if $i$ is in the set $\tset$.
In essence, the specification states that if there are $\cntvar$ distinct coupons which we have not drawn before, we need to provide exactly $n\cdot \hsum(\cntvar)$ credits to execute $\cchprog\ \arrvar\ \cntvar$.

The key part of the proof is the sampling step, where we need to distribute the $n\cdot \hsum(\cntvar)$ credits depending on the result of the $\Rand$ operation, \ie{}, we need to figure out what postcondition $\textcolor{ACMPurple}{\Phi_?}$ to use when deriving the Hoare triple $\hoare{\tc{n\cdot\hsum(\cntvar)}}{\Rand n} {\textcolor{ACMPurple}{\Phi_?}}$ with \ruleref{ht-rand-exp}.

Notice that $n\cdot \hsum(\cntvar)=\sum_{i=1}^{\cntvar}D_{i}$.
Looking back at our pen-and-paper proof, by linearity of expectations, the cost to get the $i$th coupon is only accounted for by $D_{i}$, and all $D_j$ with $i \neq j$ are independent.
Using the frame rule, we can temporarily ``set aside'' the credits needed for $D_{j}$:
\begin{align*}
  \infrule{}
  {\hoare{\tc{\tfrac{n}{\cntvar}}}{\Rand n} {\textcolor{ACMPurple}{\Phi'_?}}}
  {\hoare
  {\tc{\tfrac{n}{\cntvar}}\sep\tc{n\cdot\hsum(\cntvar-1)}}
  {\Rand n}
  {\textcolor{ACMPurple}{\Phi'_?}\sep \tc{n\cdot\hsum(\cntvar-1)}}}
\end{align*}
We proceed by the expectation-preserving composition rule and assign $\frac{n}{\cntvar}$ credit to those branches where the result of $\Rand n$ is already in $\tset$ and $0$ otherwise, \ie{}, 
\begin{align*}
  \hoare
  {\tc{\tfrac{n}{\cntvar}}}
  {\Rand n}
  {x \ldotp \left(\tc{\tfrac{n}{cnt}} \land x\in \tset\right) \lor (\tc{0} \land x\notin \tset)}
\end{align*}
as $\cost(\Rand n) + \frac{\ssize\ \tset}{n} \cdot \frac{n}{\cntvar} + \frac{n-\ssize\ \tset}{n}\cdot 0 = 1 + \frac{(n-\cntvar)}{n}\cdot \frac{n}{\cntvar} \le \frac{n}{\cntvar}$ as required.
The remainder of the proof follows by symbolic execution. 

While the coupon collector has been verified using previous program logics for expected run time analysis, such as the ERT calculus of \citet{kaminski_weakest_2016}, the structure of the proof in that calculus is quite different from the pen-and-paper analysis.
In contrast, \thetimelogic{}'s support for framing allows for reasoning about the expectation of each $D_i$, much as in the pen-and-paper proof.
Existing automated resource analysis tools, such as \texttt{eco-imp}~\cite{eco-imp}, can only provide a (non-optimal) bound for the cost of the coupon collector, whereas the specification established by \thetimelogic above is a tight bound.

\subsection{Fisher-Yates Shuffle}
\label{sec:fisher-yates-shuffle}
\newcommand{\shufflelen}{N}
\newcommand{\fyprog}{\langv{shuffle}}
\newcommand{\fyhprog}{\langv{repeatSwap}}

In this section, we use the $\costrand$ model to reason about the \emph{expected entropy cost} of the \emph{Fisher-Yates shuffle}~\cite{fisher-yates, random-permutation}.
The Fisher-Yates shuffle is a classical randomized algorithm that generates a random permutation of a sequence by repeatedly sampling an element from the sequence without replacement.
\cref{fig:fisher-yates} shows a \thelang implementation of the algorithm.
Starting from the last index $i$ of a list $\val$, we repeatedly sample an index $j$ from $\Rand i$ and swap the $i$-th and $j$-th elements of the list.
\begin{figure*}[t]
  \centering
  \begin{minipage}[t]{.45\linewidth}    
    \begin{align*}
      &\Rec \fyhprog \val~i = \\
      &\quad \If i \le 0 then \val\\
      &\quad \Else
        \begin{aligned}[t]
          &\Let j = \Rand i in \\
          &\Let \val' = \listswap~\val\ i\ j in \\
          &\fyhprog\ \val'\ (i-1)
        \end{aligned}
    \end{align*}
  \end{minipage}
  \begin{minipage}[t]{.45\linewidth}
    \begin{align*}
      \fyprog \eqdef{} & \Lam\ \val. \\
                       &\Let i = (\listlen~\val)-1 in\\
                       & \fyhprog\ \val\ i
    \end{align*}
  \end{minipage}
  
  \caption{An implementation of the Fisher-Yates shuffle.}
  \label{fig:fisher-yates}
\end{figure*}

The Fisher-Yates shuffle is unbiased in the sense that every permutation is equally likely.
Given that the input list is of length $\shufflelen$, there is a total of $N!$ possible permutations, assuming each element in the list is distinct.
Since each permutation is equally likely in an unbiased shuffle, no algorithm for unbiased shuffling can consume less than $\log_2(N!)$ entropy. 
We prove that the Fisher-Yates shuffle achieves this optimal entropy consumption in expectation, by proving an upper bound of $\log_2(N!)$ on its expected entropy cost.
In particular, we show
\begin{align*}
  \hoare{\islist(l, \val) \sep \tc{\log_2 (|l|!)}}{\fyprog\ \val}{\val' \ldotp  \Exists l'. \islist(l', v') \sep l \equiv_{p} l'}.
\end{align*}
Here $\islist(l, \val)$ is a standard representation predicate which means $\val$ corresponds to the mathematical list $l$, and $l \equiv_{p} l'$ denotes that $l$ is a permutation of $l'$.
The proof of the specification is mostly straightforward: for any natural number $N$, we have $\log_2(N!) = \sum_{i=2}^N\log_2(i) = \sum_{i=1}^{N-1} \cost (\Rand i)$, which is exactly the cost needed to pay for each $\Rand$ operation for a list of length $N$.

This example demonstrates the benefit of parameterizing by different cost models, instead of trying to use $\Tick$ to instrument a program to register its total cost.
In general, the $\log_2(i+1)$ entropy cost of calling $\Rand i $ may be an irrational real number. 
Thus, instrumenting the program with a $\Tick(\log_2(i+1))$ to incur this cost would require adding primitives to the language for computing the $\log_2$ in the argument to $\Tick$.

\subsection{Batch Sampling}
\label{sec:batch-sampling}
\newcommand{\sampleprog}{\langv{sampleThree}}
\newcommand{\resultvar}{v}
\newcommand{\flipnprog}{\langv{flipN}}
\newcommand{\batchhelperprog}{\langv{prefetch}}
\newcommand{\batchprog}{\langv{initSampler}}
\newcommand{\batchmem}{\langv{mem}}
\newcommand{\batchcnt}{\langv{cnt}}
\newcommand{\batchrem}{r}
\newcommand{\batchquot}{q}
\newcommand{\amortizedinv}{\pi}

It is a folklore result that the entropy cost of generating samples from various distributions can be reduced by generating samples in batches, instead of one at a time~\citep{DBLP:journals/corr/abs-1304-1916, DBLP:journals/tit/HanH97}.
Samples that are generated in batches can be stored in a buffer and then returned one by one, only regenerating when the buffer is exhausted.
In this section, we consider a concrete example described by \citet{DBLP:journals/corr/abs-1012-4290} of how batching can reduce per-sample entropy.
As we will see, reasoning about the buffering of batched samples is natural with amortized cost analysis.

Suppose we would like to simulate uniform sampling from a set of three elements, \eg{}, the set $\{0, 1, 2\}$ but the only probabilistic primitive that we have at our disposal is $\Rand 1$, a probabilistic  flip.
One approach is to execute $\Rand 1$ twice and decode the combined output as a 2-digit binary number.
If this number equals $3$ we repeat the process, otherwise we return the decoded number.
\begin{align*}
  \Rec \sampleprog\ \_ =\;
  &\Let \resultvar = (\Rand 1) + 2 * (\Rand 1) in\\
  &\If \resultvar < 3 then \resultvar \Else \sampleprog\ ()
\end{align*}
Since $\cost(\Rand 1) = \log_2 (1+1)= 1$, one can show that the expected entropy cost of one execution of $\sampleprog$ is $\frac{8}{3}$ by proving the following specification, which is mostly straightforward.
\begin{align*}
  \hoare{\tc{\tfrac{8}{3}}}{\sampleprog\ ()}{n \ldotp 0 \le n \leq 2}
\end{align*}

Next, we consider a batch sampling scheme shown in \cref{fig:batch-sampling}, which generates five $\Rand 2$ samples at once.
The key advantage of sampling in batches is that the \emph{amortized} expected cost will be strictly smaller than $\frac{8}{3}$, so over a long sequence of queries, it will be cheaper than $\sampleprog$.
When called, the function $\batchprog$ initializes state needed by the sampler, and then returns a batched sampling function that can be invoked to generate samples.
This returned function uses a helper function $\batchhelperprog$, which executes $\Rand 1$ eight times, intuitively corresponding to sampling a natural number strictly below $2^8=256$.
If the number is strictly below $3^5=243$, we can encode it as a $5$-digit ternary number, where each of the digits represents an independent $\Rand 2$ sample.
Otherwise, if the number is $\geq 243$, we repeat the process of executing eight $\Rand 1$ samplings until the result is below $243$.
The result of $\batchhelperprog$ is cached in the location $\batchmem$, and then each of the generated digits are returned for the next $5$ queries.
The location $\batchcnt$ stores the number of queries left before a new batch is required.
With this approach, we only incur a large cost at the start of every five queries, while the remaining four queries incur $0$ cost, as we are only reading out the corresponding result from the batched sampling.
\begin{figure*}[t]
  \centering
  \begin{minipage}[t]{.4\linewidth}
    \begin{align*}
      &\Rec \batchhelperprog \batchmem = \\
      &\quad\Let \resultvar = \flipnprog\ 8 in \\
      &\quad\If \resultvar < 243 then \\
      &\qquad \begin{aligned}
          \batchmem \gets\resultvar
        \end{aligned} \\
      &\quad\Else \batchhelperprog~m 
    \end{align*}
  \end{minipage}
  \begin{minipage}[t]{.45\linewidth}
    \begin{align*}
      \batchprog\eqdef{}
      &\Let \batchmem = \Alloc 0 in \\
      &\Let \batchcnt = \Alloc 0 in \\
      &\Lam\ \_.
        \begin{aligned}[t]
          & (\If \deref \batchcnt == 0 then \\
          & \quad
            \begin{aligned}
              &\batchhelperprog~\batchmem; \batchcnt \gets 5);
            \end{aligned} \\
          & \Let \resultvar = \deref \batchmem in \\
          & \batchcnt \gets \deref \batchcnt - 1; \\
          & \batchmem \gets \resultvar~\text{\`{}quot\`{}}~3; \\
          & \resultvar~\text{\`{}rem\`{}}~3
        \end{aligned}
    \end{align*}
  \end{minipage}  
  \caption{An implementation of batch sampling.}
  \label{fig:batch-sampling}
\end{figure*}

After an initial cost of ${4 \cdot \tfrac{256\cdot 8}{243\cdot 5}}$ to initialize the sampler, the amortized cost of each subsequent query is $\frac{256\cdot 8}{243\cdot 5}$ which is strictly smaller than $\frac{8}{3}$, the expected cost of our original $\sampleprog$ program.
We establish this claim by showing the higher-order specification
\begin{align*}
  &\hoare
  { \tc{4 \cdot \tfrac{256\cdot 8}{243\cdot 5}} }
  {\batchprog}
  { f \ldotp
    \amortizedinv \sep
    \hoare
    {\tc{\tfrac{256\cdot 8}{243\cdot 5}}\sep \amortizedinv}
    {f~()}
    {n \ldotp (0\le n<3) \sep \amortizedinv}
    }
\end{align*}
where
\begin{align*}
  \amortizedinv \eqdef{} \Exists \batchcnt, c, \batchmem, m .  \batchcnt \mapsto c\ \sep  c < 5\ \sep \batchmem \mapsto m\ \sep  m < 3^c\ \sep  \tc{(4-c) \cdot \tfrac{256\cdot 8}{243\cdot 5}}
\end{align*}
The $\amortizedinv$ assertion serves two purposes: (1) it relates the locations $\batchcnt$ and $\batchmem$ and their contents, and (2) it serves as a reserve that stores the credits provided in previous invocations to $\batchprog$.
The reserve is increased by the intermediate queries that have actual cost $0$---but non-zero amortized cost---and spent all at once for the expensive operations happening every fifth query.

We omit the details of the proof which, as in previous examples, relies on choosing how to distribute credits during the batch sampling through the expectation-preserving composition rule.
The key take-away of this example is that by treating cost as a resource in separation logic, advanced reasoning about amortized expected cost is naturally supported.

\subsection{Hash Map}
\label{sec:hash-map}
\newcommand{\computehash}{\langv{compute\_hash}}
\newcommand{\mapget}{\langv{get}}
\newcommand{\mapset}{\langv{set}}
\newcommand{\insertelem}{\langv{insert}}
\newcommand{\lookupelem}{\langv{lookup}}
\newcommand{\llinsert}{\langv{LinkedList.insert}}
\newcommand{\lllookup}{\langv{LinkedList.lookup}}
\newcommand{\hmlength}{\langv{hmLength}}
\newcommand{\hmsize}{\langv{size}}
\newcommand{\ishashmap}{\langv{isHashMap}}
\newcommand{\ishashfunction}{\langv{isHashFunction}}
\newcommand{\hmmax}{\texttt{MAX}}
\newcommand{\isamortizedhashmap}{\langv{isAmortizedHashMap}}
In this example, we show that \thetimelogic scales to reasoning about realistic data structures.

A hash map (also known as a hash table or a hash set) is a data structure widely used for implementing dictionaries, caches, and sets.
We implement a hash map that supports two operations: insertion and lookup.
Our hash map contains an array, and each array index contains a pointer to a linked list (called a bucket), which stores a list of elements.
To insert an element, the hash map uses a hash function to hash the element, and appends it to the bucket with the array index of the hash.
To perform a lookup on an element, the hash map hashes the element and traverses the bucket with the array index of the hash to determine whether it is present in the bucket.

In this example, we establish cost bounds on inserting and looking up elements in our hash map implementation. We use the $\costtick$ model, where we incur $\tc{1}$ whenever we perform an access to memory by manually adding a $\Tick 1$ operation after each dereference.

Our hash map utilizes a model of the idealized hash function (\cref{fig:hashfun}) under the uniform hash assumption~\cite{uniform_hash_assumption}, which assumes that the hash function $hf$ from a set of keys $K$ to values $V$  behaves as if, for each key $k$, the hash $hf(k)$ is randomly sampled from a uniform distribution over $V$, independently of all other keys.
We implement this model by first initializing an empty mutable map $hf$. When we want to hash a key $k$, we first check the map to determine whether there is an element under that key.
If so, we return the hash value stored in $hf(k)$.
Otherwise, we sample a value uniformly from $V=\{0, \dots, n\}$, store the value in $hf$ with key $k$, and return it.

\captionsetup{belowskip=-5pt,aboveskip=5pt}

\begin{figure*}[htb]
  \centering
  \begin{align*}
    \computehash \eqdef{}
    & \Lam\ hf\ k.\MatchML \mapget\ hf\ k with
      \Some(v) => v
      | \None => \Let v = \Rand n in \mapset\ hf\ k\ v \,;\; v
      end {}
  \end{align*}  
  \caption{An idealized hash function implementation.}
  \label{fig:hashfun}
\end{figure*}

Although the implementation of the hash function involves location accesses, we do not include $\Tick$ operations in our hash implementation.
This is because our implementation only serves to simulate the effects of a realistic hash function, which in real life, usually incurs constant  cost for each hash operation.

The two main operations supported by the hash map are \(\insertelem\) and \(\lookupelem\) as shown in \cref{fig:hashmap}.
The function $\insertelem$ hashes an element \(v\) with the hash function $hf$ to get the index \(idx\) of the bucket that \(v\) should go into and accesses the bucket \(idx\) with a dereference, incurring one tick.
Finally, \(\insertelem\) stores \(v\) at end of the linked list representing the bucket after checking for the presence of \(v\) to avoid duplication.
The function $\lookupelem$ similarly computes \(idx\) by hashing \(v\), accessing the bucket \(idx\) with a load and hence incurring one tick cost, and finally performing a lookup by traversing the linked list.
We omit the (standard) code for the linked list functions $\llinsert$ and $\lllookup$, but note that since they traverse the entire linked list, and memory dereferencing costs $1$ in our model, the functions incur $\tc{len}$ credits for each invocation where $len$ is the length of the linked list.

\begin{figure*}[htb]
  \centering
  \begin{minipage}[t]{.4\linewidth}
    \begin{align*}
      \insertelem \eqdef{} & \Lam\ h\ v.\\
                           &\Let (l, hf) = h in \\
                           & \Let idx =\computehash\ hf\ v in \\
                           & \Let b = \deref l[idx] in \\
                           & \Tick{1}; \\
                           &(l\locadd idx) \gets \llinsert\ b\ v\\
    \end{align*}
  \end{minipage}
  \begin{minipage}[t]{.4\linewidth}
    \begin{align*}
      \lookupelem \eqdef{} & \Lam\ h\ v.\\
                           &\Let (l, hf) = h in \\
                           & \Let idx =\computehash\ hf\ v in \\
                           & \Let b = \deref l[idx] in \\
                           & \Tick{1}; \\
                           &\lllookup\ b\ v
    \end{align*}
  \end{minipage}
  \caption{A hash map implementation.}
  \label{fig:hashmap}
\end{figure*}

Let us consider the  costs of $\insertelem$ and $\lookupelem$ in the case where an element has never been hashed by a hash function $hf$ before. For $\insertelem$, when we first hash the element, since it has not been hashed by $hf$ before, the idealized hash function randomly picks an index $idx$ from $V=\{0, \dots, n\}$. Let the  cost of $\insertelem$ in the case that the element is hashed to $idx$ by $hf$ be $C(idx)$. By stepping through the program, we can show that $C(idx) = 1 + \hmlength(h, idx)$, where $\hmlength(h, idx)$ is notation for the length of the linked list representing the bucket with index $idx$. %
Hence the expected cost of $\insertelem$ is $\expect[\distr]{C}$, where $\distr$ is the uniform distribution on $\{0,\ldots,n\}$. By linearity of expectations, we have $\expect[\distr]{C}=\expect[\distr]{1}+\expect[\distr]{\hmlength(h, idx)}=1+\frac{\hmsize(h)}{n+1}$ where $\hmsize(h)$ is shorthand for the total number of elements stored in the hash map (this is the sum of lengths of all the linked list buckets). The time complexity of $\lookupelem$ is exactly the same, as instead of adding an element to the end of a linked list in $\insertelem$, we still traverse the linked list comparing our element with each value stored by the linked list.

To verify the above claims, we prove that the expected cost of inserting and looking up an element not in the hash map is proportional to the size of the hash map, \cf{}, the specifications below.
In these specifications, the predicate $\ishashmap(h , m_1, m_2)$ intuitively asserts that the hash map $h$ is valid according the two mathematical maps $m_1$ and $m_2$.
The map $m_1$ models the map implementing the hash function and $m_2$ models the linked lists of each bucket of the hash map.
\begin{align*}
  &\hoareV
    {\val \notin \dom(m_{1}) \sep
    \ishashmap(h, m_1, m_2) \sep 
    \tc{1+\tfrac{\hmsize(h)}{n+1}}
    }
    {\insertelem~h~\val}
    {\_ \ldotp \Exists i . \ishashmap(h, \mapinsert{\val}{i}{m_1}, \mapinsert{i}{m_2[i] \dplus [\val]}{m_2})} \\
  &\hoareV
    {
    v\notin \dom(m_1) \sep 
    \ishashmap(h, m_1, m_2) \sep 
    \tc{1+\tfrac{\hmsize(h)}{n+1}}
    }
  {\lookupelem\ h\ v}
  {b \ldotp b = \False \sep \Exists i . \ishashmap(h, \mapinsert{v}{i}{m_1}, m_2)}
\end{align*}
In addition, we also prove that by fixing the maximum size of the hash map, we can show an amortized specification for the insertion of new elements into the hash map\footnote{There is no natural amortized specification analogue of $\lookupelem$, as the function does not change the size or structure of the hash map, and hence each invocation has the same cost. }, where the cost of insertion is a fixed constant $\rt$.
\begin{align*}
  \hoareV
  {v\notin \dom(m_1) \sep 
  \hmsize(h)<\hmmax \sep 
  \isamortizedhashmap(h, m_1, m_2) \sep 
  \tc{x}}
  {\insertelem~h~v}
  {\_ \ldotp \Exists i . \isamortizedhashmap(h, \mapinsert{v}{i}{m_1}, \mapinsert{i}{m_2[i] \dplus [v]}{m_2})}
\end{align*}

In textbook proofs of the case where an element \(v\) has previously been hashed (an thus exists in the hash table), one proves that the cost of $\insertelem$ (and $\lookupelem$) is also proportional to the size of the hash table, assuming that \(v\) is chosen uniformly across all elements stored in the hash table. This assumption is not directly expressible within \thetimelogic, and we leave this problem as future work.

\subsection{Randomized Quicksort}
\label{sec:randomized-quicksort}

\newcommand{\qs}{\langv{qSort}}
\newcommand{\partition}{\langv{partition}}

\newcommand{\lenc}{\langv{length}}
\newcommand{\qst}{\langv{t}}
\newcommand{\pivotv}{\langv{pivot}}

\newcommand{\qsl}{\langv{xs}}
\newcommand{\qsc}{\langv{cmp}}
\newcommand{\qsn}{\langv{n}}
\newcommand{\qsi}{\langv{i}}
\newcommand{\qsp}{\langv{pivot}}
\newcommand{\qsr}{\langv{rest}}
\newcommand{\qslt}{\langv{xslt}}
\newcommand{\qsgt}{\langv{xsgt}}
\newcommand{\qslts}{\langv{xslt'}}
\newcommand{\qsgts}{\langv{xsgt'}}

A common technique in randomized algorithm design is \textit{divide-and-conquer}, where
a program decomposes a problem into several ``smaller'' subproblems, often based on a random choice.
A cost analysis of a divide-and-conquer algorithm often involves first modeling the
program using a recurrence relation, and then deriving upper bounds or closed forms
for the model mathematically.
Distinguishing the \textit{modeling} and \textit{bounding} stages can be advantageous
for mechanized program verification, since the recurrence relations in the \textit{modeling} step
are typically exact (eliminating the re-engineering work from guessing a bound that is ``too loose'') and more straightforward to establish.
To show how the \textit{expectation rule} can simplify cost bound proofs for
divide-and-conquer algorithms, we verify an implementation of randomized quicksort
in \eclog.

\newcommand{\iscmp}{\langv{isPureComp}^\iota}
\newcommand{\booldecide}{\langv{bool\_decide}}

\begin{figure}[bht]
  \begin{minipage}[t]{0.33\linewidth}
    \begin{align*}
      &\Rec \qs \qsc~\qsl=\\
      & \quad \Let \qsn = \listlen\ \qsl in  \\
      & \quad \If \qsn < 1 then \qsl \Else \\
      & \qquad \Let \qsi = \Rand~(\qsn - 1) in \\
      & \qquad \Let (\qsp, \qsr) = \listremoventh \ \qsl\ \qsi in \\
      & \qquad \Let \qslt = \listfilter~(\Lam x . \Negb\ (\qsc\ \qsp\ x)) \ \qsr in \\
      & \qquad \Let \qsgt = \listfilter~(\Lam x . \qsc\ \qsp\ x) \ \qsr in \\
      & \qquad \Let (\qslts, \qsgts) = (\qs\ \qsc\ \qslt, \qs\ \qsc\ \qsgt) in \\
      & \qquad \listappend~\qslts~(\listcons~\qsp~\qsgts)
    \end{align*}
  \end{minipage}
  \caption{An implementation of Quicksort.}
  \label{fig:quicksort}
\end{figure}

\Cref{fig:quicksort} depicts a higher-order \thelang implementation of
quicksort.
The program operates on functional lists of generic values, using a function $\qsc$ to
compare them.
As is common in the analysis of sorting algorithms we will only count the costs incurred by comparisons; we can parametrically calculate the total expected cost incurred by $\qsc$
by using the $\costtick$ model since $\qs$ contains no $\Tick$ statements.
Let $R$ be a decidable, total order on a type $A$, and let $\iota$ be an injection from $A$ to $\Val$. We will use $\iota$ to specify that \thelang values are ordered according to $R$.
We define a predicate $\iscmp$ to specify that $\qsc$ correctly computes comparisons with respect to the order $R$,
incurring a bounded cost $k$:
\begin{equation*}
  \iscmp(c, R, k) : \iProp \eqdef
  \forall (x, y : A),\;
  \hoare
    { \tc{k} }
    {c\ (\iota\ x)\ (\iota\ y)}
    {b \ldotp b = R\ x\ y}
\end{equation*}

Given a constant $m \geq 0$ define $\qst_{m}(n)$ by the recurrence relation
\begin{equation*}
  \qst_{m}(n) \eqdef
  \begin{cases}
    0 & n = 0 \\
    mn + \frac{2}{n} \sum_{i=0}^{n-1} \qst_{m}(i) & n > 0
  \end{cases}
\end{equation*}
A standard mathematical derivation (which we will omit here, but which is
formalized in our Coq development) shows that $\qst_{m}(n)$ is bounded above by $(n+1) \sum_{i=1}^{n}\frac{2m}{i+1}$.
It is well known that the harmonic series is $\mathcal{O}(\log{} n)$ so establishing
a cost bound of $\qst_{m}(n)$ for any $m$ is enough to obtain the familiar
$\mathcal{O}(n\log{}n)$ asymptotic result.

We can now move on to specifying and verifying our implementation of $\qs$.
Let $L$ be a list of values of type $A$ without duplicates, and let $v$ be a value.
By applying $\iota$ to each element, define the predicate $\islist(L, v)$
to hold when $v$ is a value-level reflection of $L$.
We then seek to prove that
\begin{equation*}
  \hoareV
  {\tc{\qst_{2k}(|L|)} \sep 
     \islist(L, v) \sep 
     \listnodup(L) \sep 
     \iscmp(\qsc, R, k)
   }
  {\qs~\qsc~v}
  { v' \ldotp \Exists L' .
    L \equiv_{p} L' \sep 
    \listsorted(L') \sep 
    \islist(L', v')
  }
\end{equation*}

The predicates in the postcondition 
state the functional correctness of quicksort, namely that the returned list is a sorted permutation of the input.
Since the functional correctness is irrelevant for the runtime bound we will elide that
part of the argument here, and just focus on explaining the cost accounting.
We proceed by induction using \rref{ht-rec}.
Set $N = |L|$ and, after eliminating the $N=0$ edge case, split $\tc{\qst_{2k}(N)}$
into $\tc{kN + kN}$ and $\tc{\frac{2}{N} \sum_{i=0}^{N-1} \qst_{2k}(i)}$.
We will spend the former credit to execute the comparisons in the two
$\listfilter$ statements, as in our implementation their cost will be the same regardless of the
pivot element selected by $\Rand$.
We will eventually use the latter credit to pay for the induction hypothesis during the recursive calls to $\qs$.

\newcommand{\ixrk}{\langv{ix\_rk}}
\newcommand{\rkrev}{\langv{rk\_rev}}
\newcommand{\qscmp}{d}

When the sampled pivot is close to the largest or smallest element in the list the latter credit is not be enough to pay for the recursive calls (conversely, we have a credit excess when the pivot is near the median).
To ensure that we can apply the induction hypothesis in all
cases we must redistribute the credits at the statement $\Rand~(n-1)$ based on the
rank of the pivot.
Define
\begin{align*}
  \ixrk_{L}(i) &\eqdef |\{j \in L \;\vert\; j < L[i]\}| \\
  \rkrev_{L}(i) &\eqdef N - 1 - i
\end{align*}
The function $\ixrk_{L}$ converts an index $i$ in $L$ into the rank of $L[i]$,
and the function $\rkrev_{L}$ relates the number of elements greater than the pivot
to the rank of the pivot itself.
Crucially both functions are bijections on $\{0, \ldots, N - 1 \}$, the sample space of $\Rand~(N - 1)$, so post-composing a sum over that space with $\ixrk_{L}$ or $\rkrev_{L}$ will not change its value.
Using our bijections, we can define an credit distribution function.
\begin{equation*}
  \qscmp_{L}(i) = (\qst_{2k} \circ \ixrk_{L})(i) + (\qst_{2k} \circ \rkrev_{L} \circ \ixrk_{L})(i)
\end{equation*}

We can show that the mean of $\qscmp_{L}$ is exactly the credit remaining after
framing out $\tc{kN + kN}$, meaning we can apply $\ruleref{ht-rand-exp}$ using $\qscmp_{L}$
at $\Rand\ (n - 1)$:
\begin{align*}
  \frac{1}{N} \sum_{i=0}^{N-1} \qscmp_{L}(i)
  & = \frac{1}{N} \sum_{i=0}^{N-1} (\qst_{2k} \circ \ixrk_{L})(i) + \frac{1}{N} \sum_{i=0}^{N-1} (\qst_{2k} \circ \rkrev_{L} \circ \ixrk_{L})(i) \\
  & = \frac{1}{N} \sum_{i=0}^{N-1} \qst_{2k}(i) + \frac{1}{N} \sum_{i=0}^{N-1} \qst_{2k} (i) = \frac{2}{N} \sum_{i=0}^{N-1} \qst_{2k}(i)
\end{align*}

Therefore, when $\Rand\ (n - 1)$ selects the pivot index $i$, we can use
$\tc{kN + kN + \qscmp_{L}(i)}$ to verify the continuation.
Since $\ixrk_{L}(i) = \lenc\ lt$ and $(\rkrev_{L} \circ \ixrk_{L})(i) = \lenc\ gt$
this is precisely enough credit to filter the lists and then apply the induction hypothesis,
completing the proof.

\paragraph{Verifying the Expected Entropy Cost for Quicksort}

\newcommand{\qsent}{\langv{e}}

Using \eclog to establish a recurrence relation for a cost bound and then separately analyzing the recurrence leads to more modular proofs and proof re-use.
For example, when we change cost models, often the overall \emph{structure} of a recurrence relation will not change, and so the same proof can be adapted to establish a recurrence under this different model, even if analyzing the asymptotics of the recurrence or computing a closed form would be quite different.
To illustrate this, we verify a recurrence for the expected amount of entropy consumed by the
random choices in $\qs$.
Define a recurrence relation $\qsent(n)$ which represents the entropy required
to sort a list of length $n$:
\begin{equation*}
  \qsent(n) \eqdef
  \begin{cases}
    0 & n = 0 \\
    \log_{2}(n) + \frac{2}{n} \sum_{i=0}^{n-1} \qsent(i) & n > 0
  \end{cases}
\end{equation*}
Given a comparison function with zero entropy cost, an argument directly analogous
to the time complexity proof shows that $\tc{\qsent(n)}$ suffices to execute $\qs$
on lists of length $n$ under the $\costrand$ model.
The proofs primarily differ by where they spend credit: the $\costtick$ proof
spends credit inside $\listfilter$, whereas the $\costrand$ proof spends
credit to execute $\Rand$.
The details of the expectation-preserving transfer of credits across branches and other steps of program execution are otherwise nearly identical, even reusing much of their lemmas.
It is known that $\qsent(n)$ is asymptotically linear~\citep[p. 89-91]{cover2006elements}, 
but solving this recurrence or computing the asymptotics of $\qsent(n)$ is quite different from the analysis of $\qst_{m}(n)$, the recurrence for the cost of comparisons.
Thus, if the logic had required solving or bounding the recurrence at the same time as we analyzed the program, we would not be able to re-use as much of the proof.

\subsection{Randomized Heaps}
\label{sec:randomized-heaps}

\newcommand{\hinsert}{\langv{insert}}
\newcommand{\hremovemin}{\langv{remove}}
\newcommand{\hmeld}{\langv{meld}}
\newcommand{\hc}{\langv{cmp}}
\newcommand{\hv}{\val}
\newcommand{\hr}{\langv{r}}
\newcommand{\hh}{\langv{h}}
\newcommand{\hhl}{\langv{hl}}
\newcommand{\hhr}{\langv{hr}}
\newcommand{\symhmin}{\langv{min}}
\newcommand{\symhmax}{\langv{max}}

\begin{figure}[t]
  \begin{minipage}[t]{0.33\linewidth}
    \begin{align*}
      \hmeld \eqdef\; 
      & \Rec \hmeld\ \hc\  \hv_{1}\ \hv_{2} =\\
      & \Match \hv_{1} with \None => \hv_{2} | \Some \hh_{1} => \\
      & \Match \hv_{2} with \None => \hv_{1} | \Some \hh_{2} => \\
      & \Let ((\hr_{\symhmin}, \; (\hhl_{\symhmin}, \; \hhr_{\symhmin})), \; \hh_{\symhmax}) = \\
      & \quad (\If \hc\ (\Fst \hh_{1})\ (\Fst \hh_{2}) then (\hh_{1}, \hh_{2}) \Else (\hh_{2}, \hh_{1}) ) in \\
      & \If \Flip \\
      & \quad then \Some~(\hr_{\symhmin}, (\hmeld\ \hc\ \hhl_{\symhmin}\ \Some(\hh_{\symhmax}),\; \hhr_{\symhmin})) \\
      & \quad \Else \Some~(\hr_{\symhmin}, (\hhl_{\symhmin},\; \hmeld\ \hc\ \hhr_{\symhmin}\ \Some(\hh_{\symhmax}))) \\
      & end end \\
      \hinsert \eqdef \;
      & \Lam \hc, \hr, \hv . (\hr \leftarrow \hmeld\ \hc\ (\deref \hr) \ \Some(\hv, (\None, \None))) \\
      \hremovemin \eqdef \;
      & \Lam \hc, \hr . \\
      & \Match \deref \hr with \\
      & \;\; \None => \None \\
      & | \Some~(\hv,\; (\hhl,\; \hhr)) => (\hv \leftarrow \hmeld\ \hc \ \hhl \ \hhr); \; \Some\ \hv \\
      & end
    \end{align*}
  \end{minipage}
  \caption{An implementation of randomized meldable heaps.}
  \label{fig:meldheap}
\end{figure}

A \textit{randomized meldable heap}~\citep{DBLP:conf/sofsem/GambinM98} is an implementation of a binary heap, which use randomization
to perform $\hinsert$ and $\hremovemin$ using an expected $\mathcal{O}(\log(n))$ comparisons.
The key to this performance is the $\hmeld$ operation, which recursively
combines two meldable heaps by melding the heap with greater root into a
random child of the heap with lesser root.
Choosing a child of the lesser heap at random ensures the \textit{expected} size
of a heap in the recursive call to $\hmeld$ is halved, regardless of how unbalanced the children may be.
This is essential to obtaining the expected cost bound.
Below, we verify an implementation of meldable heaps against a modular
heap specification, using expectation preserving composition to obtain expected
cost bounds.

\newcommand{\isComparator}{\mathsf{isComp}}
\newcommand{\cmpkey}{\mathsf{hasKey}}

\begin{figure}[t]
  \begin{align*}
    \isComparator(K, \textlang{cmp}, \cmpkey) \eqdef{}
    & \Exists\, R : K \to K \to \bool, \rt : \nnreal. \mathsf{PreOrder}(R) \land \mathsf{Total}(R) \land{} \\
    & \quad
      \hoareV{\cmpkey(k_{1}, \val_{2}) \sep \cmpkey(k_{2}, \val_{2}) \sep \tc{\rt}}
      {\textlang{cmp}~\val_{1}~\val_{2}}
      { b \ldotp b = R(k_{1}, k_{2}) \sep \cmpkey(k_{1}, \val_{2}) \sep \cmpkey(k_{2}, \val_{2})}
  \end{align*}
  \caption{A specification for an abstract comparator.}
  \label{fig:compspec}
\end{figure}

\newcommand{\isminheap}{\mathsf{isHeap}}
\newcommand{\mnew}{\textlang{new}}
\newcommand{\minsert}{\textlang{insert}}
\newcommand{\mremove}{\textlang{remove}}

\begin{figure}[t]
\begin{align*}
  & \isComparator(K, \textlang{cmp}, \cmpkey) \Rightarrow \\
  & \Exists\, \isminheap : \List(K) \to \Val \to \iProp, \Rt_{i}, \Rt_{r} : \nat \to \nnreal . \\
  & \phantom{\land{}}\;\;\, (\All n, m . n \leq m \Ra \Rt_{i}(n) \leq \Rt_{i}(m))
    \land (\All n, m . n \leq m \Ra \Rt_{r}(n) \leq \Rt_{r}(m)) \\
  & \land{} \; \hoare{\TRUE}{\mnew~\TT}{\val \ldotp \isminheap([], v)} \\
  & \land{} \;
    \hoare
    {\isminheap(l, v) \sep \cmpkey(k, \valB) \sep \tc{\Rt_{i}(|l|)}}
    {\minsert~\val~\valB}
    { \_ \ldotp \Exists l' . \isminheap(l', \val) \sep l \equiv_{p} (k :: l')} \\
  & \land{} \;
    \hoareV[t]
    {\isminheap(l, v) \sep \tc{\Rt_{r}(|l|}}
    {\mremove~\val}
    { \valB \ldotp
    \begin{array}{l}
      \left(\valB = \None \sep l = [] \sep \isminheap([], \val)\right) \\
      {} \lor
      \left(\Exists u, k, l' . \valB = \Some~u \sep l \equiv_{p} (k :: l')
        \sep \mathsf{min}(k, l) \sep \cmpkey(k, u) \sep \isminheap(l', \val)\right)
    \end{array}}
\end{align*}

  \caption{An abstract specification for a min-heap.}
  \label{fig:heapspec}
\end{figure}

\cref{fig:meldheap} depicts a \thelang implementation of meldable heaps, and
\cref{fig:heapspec} outlines our target specification.
Our heap specification is parameterized by a \textit{comparator}, a \thelang program
which we specify (in \cref{fig:compspec}) to implement a total preorder using a bounded
cost $x$.
The comparator specification also includes a resourceful representation predicate
$\cmpkey : K \to \Val \to \iProp$, connecting comparable elements of type $K$ to a particular
\thelang values.\footnote{We remark that this differs from the representation predicate in \cref{sec:randomized-quicksort}, since our implementation of $\listfilter$ does not respect linearity, and $\qs$ is specified against a partial order, rather than a preorder.}
In \cref{sec:kway-merge} we will apply the heap specification with a concrete
comparator instance, demonstrating the modularity of cost bounds proven in \eclog.

\newcommand{\comparator}{\color{red}{COMPARATOR}}
\newcommand{\cmphaskey}{\langv{key}_\hc}

\newcommand{\binarytree}{\langv{BinaryTree}}
\newcommand{\isHeapB}{\langv{heap}}
\newcommand{\isHeap}{\textcolor{red}{ISHEAP}}
\newcommand{\reprbt}{\langv{isBinaryTree}}
\newcommand{\treelist}{\langv{treeToList}}
\newcommand{\ismeldheapval}{\langv{isMeldHeapVal}}
\newcommand{\ismeldheapref}{\langv{isMeldHeapRef}}

For the remainder of this section, we assume that $A$, $\hc$, and $\cmphaskey$
satisfy the predicate $\isComparator(A, \hc, \cmphaskey)$ with some cost bound $k \geq 0$ and total preorder $R$.

Define a meta-level type $\binarytree$ of binary trees at type $A$, and a
function $\treelist$ which produces a list of the elements stored in a binary tree
in an unspecified order.
We represent a binary tree in \thelang as either $\None$ when the tree is empty, or $\Some~(v, (l, r))$ when the tree is a node with value $v$ and children $l$ and $r$.
By lifting $\cmphaskey$ to the nodes of this structure we obtain a representation
predicate for binary trees, denoted $\reprbt : \binarytree \to \Val \to \iProp$.

We define the predicate $\isHeapB$, which holds when a meta-level binary tree is
heap-ordered with respect to $R$.
Together, we can define a specification $\ismeldheapval$ for \thelang-level heaps in terms of their
list of elements $L$: the value must represent a binary tree, which is a heap with respect to $R$, and whose elements are a permutation of $L$.
\begin{align*}
  \ismeldheapval(\hv, L) \eqdef \Exists b . \reprbt(b, \hv) \sep
  \isHeapB\ b \sep L \equiv_{p} \treelist(b)
\end{align*}
We can also lift this specification to references, which we use as our instance
for $\isminheap$ in \cref{fig:heapspec}.
\begin{align*}
  \ismeldheapref(\ell, L) \eqdef \Exists h . \ell \mapsto h \sep \ismeldheapval(h, L)
\end{align*}

Under this definition it is straightforward to
implement and verify $\mnew$, which allocates a reference to a new empty heap.

\newcommand{\tcmeld}{t^\langv{meld}}

Moving on, we will prove a specification for $\hmeld$ in \eclog.
We define the cost associated to melding a heap of size $n$ with the following function:
\begin{equation}
  \tcmeld(n)\eqdef
  \begin{cases}
    0 & n = 0 \\
    2k(1 + \log_{2}(n)) & n > 0
  \end{cases}
\end{equation}

Using $\vert \cdot \vert$ to denote both the length of a list and the size of a heap, we define
a Hoare triple for $\hmeld$ which encompasses both the cost bound and the
functional correctness.
\begin{align*}
  \hoareV
  { \ismeldheapval(\hv_{1}, L_{1}) \sep
    \ismeldheapval(\hv_{2}, L_{2}) \sep
    \tc{k + \tcmeld(|L_{1}|) + \tcmeld(|L_{2}|)}
  }
  {\hmeld~\hv_{1}~\hv_{2} }
  {\hv' \ldotp \Exists L .
      \ismeldheapval(\hv', L) \sep
      L \equiv_p L_{1} \dplus L_{2}}
\end{align*}

We proceed by \rref{ht-rec}.
After handling base cases, the body of $\hmeld$ compares the roots of the
two heaps using $\hc$, incurring a cost of $k$ credits.
We denote the minumum- and maxumum-rooted heap $\hh_{\symhmin}$ and $\hh_{\symhmax}$ respectively, by
heap ordering we know that no elements of $\hh_{\symhmax}$ are lesser than the root $\hr_{\symhmin}$ of $\hh_{\symhmin}$.
At this point we own $\tcmeld(|\hh_{\symhmax}|)$ and $\tcmeld(1 + |\hhl_{\symhmin}| + |\hhr_{\symhmin}|)$ credits.
One can observe that depending on the balance of $\hh_{\symhmin}$ this may
not be enough credits to directly apply the inductive hypothesis;
as we alluded to in the introduction, we will obtain the correct number of credits
using an expectation-preserving composition.

\newcommand{\mhdistr}{\langv{tc}^{\textrm{dist}}}

\begin{equation*}
  \mhdistr(s) \eqdef
  \begin{cases}
    k + \tcmeld(|\hhl_{\symhmin}|) & s = 0 \\
    k + \tcmeld(|\hhr_{\symhmax}|) & s = 1
  \end{cases}
\end{equation*}

In our development we show that $\tcmeld(|\hh_{\symhmin}|)$ is enough credit to apply the expectation rule with distribution function $\mhdistr$, in particular:
\begin{align*}
  [k + \tcmeld(|\hhl_{\symhmin}|) + (k + \tcmeld(|\hhr_{\symhmin}|))] / 2
  & \leq \tcmeld(1 + |\hhl_{\symhmin}| + |\hhr_{\symhmin}|)
\end{align*}

Therefore by framing, when $\Rand\ 1$ samples $s$ we will have total credit balance $\tcmeld(|\hh_{\symhmax}| + \mhdistr(s))$.
Suppose without loss of generality that $\Rand 1$ samples $0$.
Spending $\tcmeld(|h_{\symhmax}|) + k + \tcmeld(|hl_{\symhmin}|)$ credits, we can apply the inductive
hypothesis to evaluate the recursive call, $\hmeld\ \hc\ \hhl_{\symhmin}\ \Some(\hh_{\symhmax})$.
The resulting value is certainly a heap with no elements greater than $\hr$, as
none of its inputs were, so our return value also does not violate the heap order.
Furthermore, the returned heap has precisely the same elements as $L_{1} \dplus L_{2}$
up to permutation.
This establishes the postcondition.

Because $\hinsert$ and $\hremovemin$ both make one call to $\hmeld$ it is straightforward to prove the remaining specifications in \cref{fig:heapspec} from the specification for $\hmeld$, with the insertion and removal cost bounds $\Rt_{i} (N) = k + \tcmeld(N)$ and $\Rt_{r}(N) = k + 2 \tcmeld(N)$.

\citet{DBLP:conf/cav/LeutgebMZ22} describe a tool that can derive the cost bound for melding automatically.
However, the tool's input language is a first-order functional language that does not have mutable state. 
Thus, it cannot handle the parameterized comparator we consider here, which is essential for an example that makes use of the heaps that we describe next in \cref{sec:kway-merge}.

\subsection{K-way Merge}
\label{sec:kway-merge}
\newcommand{\kwayrremove}{\mathsf{repeatRemove}}
\newcommand{\kwaymerge}{\mathsf{merge}}

The \emph{k-way merge} problem \cite{cormen3rd} consists of merging $k$ sorted lists to produce a single sorted list with the same elements.
One way to solve the problem is by maintaining a min-heap of the $k$ lists, each keyed by the head of the (sorted) list \cite{DBLP:books/daglib/0067352}.
The procedure then repeatedly extracts the minimal element from the heap, adding the head element to the output buffer, and reinserting the remaining list into the heap.
\cref{fig:kway-merge} shows a \thelang implementation of the algorithm that uses \emph{some} min-heap implementation with operations $\mnew$, $\minsert$, and $\mremove$.
To showcase the modularity of \thetimelogic, we verify the k-way merge implementation in isolation, assuming \emph{only} the existence of an implementation that satisfies the abstract min-heap specification shown in \cref{fig:heapspec}.
In the end, we compose this specification with the meldable heap specification from \cref{sec:randomized-heaps} and obtain the expected $\mathcal{O}(n\log{k})$ cost bound in a fully modular way.

\begin{figure}[t]
  \begin{minipage}[t]{0.45\linewidth}
    \begin{align*}
      &\Rec \kwayrremove h~r = \\
      &\quad \MatchMLnarrow { \mremove~h } with
        \Some zs =>
        {\begin{aligned}[t]
           \MatchMLnarrow { \listhead~zs } with
           \Some z =>
           {\begin{aligned}[t]
              &r \gets \listcons~z~(\deref r) \\
              &\minsert~h~(\listtail~zs); \\
              &\kwayrremove~h~r
            \end{aligned}} |
           \None => \kwayrremove~h~r
           end {}
         \end{aligned}} |
        \None => ()
        end {}
    \end{align*}  
  \end{minipage}
  \hfill
  \begin{minipage}[t]{0.5\linewidth}
    \begin{align*}  
      \kwaymerge \eqdef{}
      \Lam zss .
      & \Let h = \mnew~\TT in \\
      & \listiter~(\minsert~h)~zss; \\
      & \Let r = \Alloc~(\listnew~\TT) in \\
      & \kwayrremove~h~r; \\
      & \listrev~\deref r
    \end{align*}
  \end{minipage}
  \caption{A implementation of k-way merge.}
  \label{fig:kway-merge}
\end{figure}

To verify $\kwaymerge$, we instantiate the $\isComparator$ interface from \cref{fig:compspec} with a comparator that compares lists by comparing their head elements.
The comparator induces cost 1 per invocation by including a $\Tick 1$ statement.
Using the abstract min-heap specification, we prove
\begin{align*}
  \hoareV
  { \islist(zss, \val) \sep (\All zs \in zss . \listsorted_{\leq}(zs)) \sep \tc{\rt_{\kwaymerge}} }
  { \kwaymerge~\val }
  { \valB . \Exists zs . \islist(zs, \valB) \sep zs \equiv_{p} \listconcat(zss) \sep \listsorted_{\leq}(zs) }
\end{align*}
where $\rt_{\kwaymerge} = (k + n) \cdot \Rt_{i}(k) + \Rt_{r}(0)$ denoting $|\listconcat(zss)|$ by $n$ and $|zss|$ by $k$.
Note that $\Rt_{i}$ and $\Rt_{r}$ are abstract cost functions unknown to the specification of $\kwaymerge$.
The proof of the specification contains no probabilistic reasoning but manages the cost as an abstract entity required as a precondition for the heap operations.
The k-way merge proof showcases the extensive modularity supported by \thetimelogic both in terms of logical abstractions, cost, and implementations.

\section{The Semantic Model and the Soundness Theorem}
\label{sec:model-soundness}

The semantic model of \eclog{} is simplified by working with \emph{weakest preconditions} instead of Hoare triples directly. The two notions are well-known to be equivalent, and we use a standard definition~\cite[\S6]{irisjournal} of Hoare triples in terms of weakest preconditions:
\[
  \hoare P \expr Q \eqdef{} \always(P \wand \wpre \expr Q)
\]

With this definition in hand, we will now turn our attention to the development of the weakest precondition and then prove adequacy of \eclog for cost and functional correctness.

\subsection{The Weakest Precondition, Cost Credits, and the Expected Cost Modality}
\label{sec:wp}

The weakest precondition associated to an expression $\expr_1$ for a postcondition $\Phi$ is defined as a guarded recursive fixed point.\footnote{%
We have greyed out the ``masks'' \(\ghostcode{\mask}\) and ``update modalities'' \(\ghostcode{\pvs[\mask]}\),
since they are (1) orthogonal to the issues discussed here and (2) our use of them in the definition of the weakest precondition is standard; see, \eg, \citet{irisjournal} for details.}
The \defemph{cost interpretation} $\tc \rt$ and $\tcauth \rt$ and the \defemph{expected cost modality} $\ecm$ constitute the main novelty of our model.
\begin{align}
  \nonumber
  \wpre{\expr_1}[\ghostcode{\mask}]{\Phi}
  \eqdef{}
  &(\expr_1 \in \Val \mathrel{\land} \ghostcode{\pvs[\mask]} \Phi(\expr_1))\\
  \nonumber
  \lor{}\,
  &(\expr_1 \not\in \Val \mathrel{\land} \All \state_1, \rt_1 . \stateinterp(\state_1) \sep \tcauth{\rt_1} \wand \ghostcode{\pvs[\mask][\emptyset]}\\
  \nonumber
  & \quad
    \ecm
    (\underbrace{\expr_1,\ \state_1}_{\cfg_1},\ \rt_1,
    \ \underbrace{(\Lam \, \expr_2, \state_2, \rt_2
    \ . \later \ghostcode{\pvs[\emptyset][\mask]} (\stateinterp(\state_2) \sep \tcauth{\rt_2} \sep \wpre{\expr_2}[\ghostcode{\mask}]{\Phi}))}_{Z})
  \\[1em]
  \nonumber
  \text{where\; }
  \ecm (\cfg_1, \rt_1, Z) \eqdef{}
  & \Exists\ (\Rt_2 : \Cfg \ra \nnreal)\ .
  \\
  \label{eq:ecm-boring}
  &\qquad \red (\cfg_1)
    \sep
    \Exists r . \All \cfg_2 . \Rt_2(\cfg_2) \leq r
    \sep
  \\
  \label{eq:ecm-exp}
  &\qquad
    \cost(\cfg_1) + \textstyle\sum_{\cfg_2 \in \Conf} \stepdistr(\cfg_1)(\cfg_2) \cdot \Rt_2(\cfg_2) \leq \rt_1
    \sep
  \\
  \label{eq:ecm-rest}
  &\qquad
    \All \cfg_2 .
    \stepdistr(\cfg_1)(\cfg_2) > 0 \wand
    Z(\cfg_2, \Rt_2(\cfg_2))
\end{align}
If $\expr_1$ is a value and $\Phi(\expr_1)$ holds, then we can prove the first branch of the disjunction.
Otherwise, under the assumption that the \defemph{state interpretation} $\stateinterp(\state_1)$ and the \defemph{cost interpretation} $\tcauth {\rt_1}$ hold, we have to prove that the \defemph{expected cost modality} $\ecm$ holds for the current configuration $(\expr_1, \state_1)$ and cost $\rt_1$.
The state interpretation ties the physical state of the program, which evolves according to the operational semantics, to the logical presentation of the heap, \ie,  the $\loc \mapsto \val$ proposition.
The last argument to $\ecm$, let us call it $Z$, contains a guarded recursive occurrence of the weakest precondition under a later modality $\later$, which ensures existence of the guarded fixpoint.

\paragraph{The cost credit resource algebra}
\label{sec:ra}
The cost interpretation $\tcauth{\rt_1}$ connects the cost budget $\tc{\rt}$ that a user of the logic manipulates to the cost incurred by the reductions of the program $\expr_1$ according to the operational semantics.
This ensures that the cost credits evolve in a way that matches the cost of the program.
We achieve this connection by defining a new (unital) resource algebra $\Auth(\nnreal, +)$ of non-negative real numbers, where $\tcauth {\rt}$ denotes the authoritative view, and $\tc \rt$ stands for ownership of a fragmental view.
The construction comes with an \emph{agreement} rule $\tc{\rt_1} \sep \tcauth{\rt_2} \vdash \rt_1 \preceq \rt_2$, a rule that allows \emph{spending} credits by updating $\tc{\rt_1} \sep \tcauth{\rt_1 + \rt_{2}}$ to $\tc{\rt_{2}}$, and a rule for \emph{acquiring} credits by updating $\tcauth{\rt_1}$ to $\tc{\rt_2} \sep \tcauth{\rt_1 + \rt_{2}}$.
The cost splitting rule from \cref{sec:base-logic} follows similarly from the laws for $(\nnreal, +)$.

\paragraph{The expected cost modality}
\label{sec:ecm}
The flexibility of \eclog to split a cost budget so long as it is preserved in expectation stems from the way the cost credits are handled in the expected cost modality $\ecm$.
The existential quantifier over $\Rt_2$ allows us to specify a family of costs indexed by configurations.
The first two clauses \eqref{eq:ecm-boring} of $\ecm$ serve bureaucratic purposes: $\red(\cfg_1)$ enforces that $\cfg_1$ is reducible (hence no verified program will get ``stuck''), and $r$, an upper bound to $\Rt_2$, is required for the existence of the sum in the following line.

The heart of the matter is the weighted sum \eqref{eq:ecm-exp}.
It requires us to distribute the currently available cost budget $\rt_1$ between the cost for stepping the current configuration $\cost(\cfg_1)$ and the cost for the remainder of the program $\Rt_2(\cfg_2)$.
Crucially, the cost for $\cfg_2$ is \emph{weighted} by the probability of stepping to it from $\cfg_1$.\footnote{The sum thus only ``counts'' configurations that are reachable from $\cfg_1$, as the probability would be 0 otherwise.}

Finally, by \eqref{eq:ecm-rest}, one has to prove that for any configuration $\cfg_2$ reachable in one step, the predicate $Z$ holds with the corresponding cost $\Rt_2(\cfg_2)$. Since we applied $\ecm$ in the definition of the weakest precondition with $Z(\expr_2,\state_2,\rt_2)=~\later{}(\stateinterp(\state_2) \sep \tcauth{\rt_2} \sep \wpre {\expr_2} \Phi)$, this amounts to carrying on proving the weakest precondition of the new expression $\expr_2$ with the updated physical state $\state_2$ and cost budget $\rt_2 = \Rt_2(\expr_2,\state_2)$.

\subsection{Soundness for Cost and Correctness}
\label{sec:soundness}

With the definition of the weakest precondition in hand, we can now prove that \eclog is a sound method for establishing cost bounds and functional correctness of \thelang programs.

\begin{theorem}[Adequacy]
  Let $\rt$ be a non-negative real number and let $\varphi$ be a predicate on values.
  If~$\tc \rt \vdash \wpre \expr \varphi$ then for any state $\state$,
  \begin{enumerate}
  \item \label{thm:wp-adequacy-cost} $\ec{\expr,\state} \leq \rt$, and
  \item \label{thm:wp-adequacy-correctness} $\All \val \in \Val\, . \exec(\expr,\state)(\val) > 0 \implies \varphi(\val)$.
  \end{enumerate}
\end{theorem}
\noindent \Cref{thm:adequacy-hoare} then follows directly from the definition of Hoare triples.
\begin{proof}
  We will focus on the salient steps of the proof of \eqref{thm:wp-adequacy-cost}; \eqref{thm:wp-adequacy-correctness} is similar and details can be found in the accompanying \rocq formalization.
  Since $\ec{\expr,\state}$ is defined as the supremum of $\ec[n]{\expr,\state}$ over $n \in \omega$, it suffices to show  $\ec[n]{\expr,\state} \leq \rt$ for all $n$.
  This in turn follows the standard soundness theorem of the Iris step-indexed logic~\cite{irisjournal} if we can derive $\vdash \later^n\ \ec[n]{\expr,\state} \leq \rt$  in Iris.

  We proceed by induction on $n$. The base case, $\vdash \ec[0]{\expr,\state} \leq \rt$, is trivial since $\ec[0]{\expr,\state} = 0$.
  Consider the inductive step with $n = m+1$ and $\expr \not\in \Val$.
  In this case, the assumption %
  amounts to
  \begin{equation}
    \label{eq:hwp}
    \ecm(\expr,\state,\rt, (\Lam \expr_2, \state_2, \rt_2 .
    \later{}\ (\stateinterp(\state_2) \sep \tcauth{\rt_2}
    \sep \wpre {\expr_2} \varphi))).
  \end{equation}
  We replace the recursive occurrence of $\wpre {\expr_2} \varphi$ with the expected cost of $(\expr_2,\state_2)$ by applying the induction hypothesis \eqref{eq:ih} under the cost modality in \eqref{eq:hwp}.
  \begin{equation}
    \label{eq:ih}
    \All \expr_2, \state_2, \rt_2 . \stateinterp(\state_2) \sep \tcauth{\rt_2}
    \sep \wpre {\expr_2} \varphi \wand \later{}^m\ \ec[m]{\expr_2,\state_2} \leq \rt_2
  \end{equation}
  We are thus left to show the following entailment:
  \[\ecm(\expr, \state, \rt, (\Lam \expr_2, \state_2, \rt_2 . \later\ (\textstyle\later^{m}\ \ec[m]{\expr_2, \state_2} \leq \rt_2))) \vdash
    \later^{m+1}\ \ec[m+1]{\expr, \state} \leq \rt
  \]
  This fact in turn is proven by a careful rearrangement of the weighted sums arising from the definitions of $\ecm$ and $\EC$.
\end{proof}

We remark here that we have also proven that a program with finite expected cost for the $\costall$ model terminates almost-surely (\ie, $\execTerm{(\expr,\state)}=1$). This is of course no surprise, but rather serves as a check that our definitions relating to costs have the usual expected properties.

\section{Related and Future Work}
\label{sec:related-future-work}

\subsection{Related Work}
\label{sec:related-work}

\paragraph{Logics for expected runtime}

\newcommand{\ert}[2]{\textsf{ert}(#1)\{#2\}}

Formal methods to reason about the expected cost of first-order programs have
seen a surge in recent years. For instance, the line of work by~\citet{kaminski_weakest_2016}
on expected runtime transformers generalize the weakest pre-expectation transformer
by~\citet{DBLP:series/mcs/McIverM05} to reason about the expected running time of programs.
This work has multiple extensions including a recent one to reason about amortized expected
cost~\cite{DBLP:journals/pacmpl/BatzKKMV23}, which makes use of separation logic assertions by adapting the approach of quantitative separation logic~(QSL)~\citep{quantitativesl}.
In these logics, assertions are interpreted as functions of type $\textlog{State} \rightarrow \mathbb{R}^{\infty}_{\geq 0}$.
Program reasoning is done using an expected-runtime transformer, $\textsf{ert}$, which is defined
in such a way that $\ert{c}{\mathbf{0}}$ is a function that maps a state $\sigma$ to the expected
running time of $c$ starting from $\sigma$.
The analogue of \thetimelogic' \ruleref{ht-rand-exp} arises from the definition of $\textsf{ert}$ for the command $c_1 [p] c_2$, which executes $c_1$ with probability $p$ and $c_2$ with probability $(1 - p)$.
This command's $\textsf{ert}$ is defined by $\ert{c_1 [p] c_2}{f} = p \cdot \ert{c_1}{f} + (1 - p) \cdot \ert{c_2}{f}$, taking the weighted average of the $\textsf{ert}$ for each command.
In contrast to these logics, in \thetimelogic, assertions are not given a quantitative interpretation and have the standard resource algebra semantics found in Iris.
A benefit of our approach is that for reasoning about non-cost-related parts of program correctness, \thetimelogic inherits all of Iris' standard, expressive reasoning rules.
The eHL logic~\cite{ehl} supports proving bounds on expected values of program states for first-order programs. Under the assumption that these programs terminate, a cost analysis can be encoded by instrumenting programs with a step counter. By integrating with a relational logic, eHL supports replacing a program with a simplified model if they compute the same costs.

Amortized reasoning about expected values has also been developed in several probabilistic extensions of Automatic Amortized Resource Analysis (AARA)~\citep{DBLP:conf/popl/HofmannJ03, DBLP:journals/mscs/HoffmannJ22}.
In the AARA approach, the type system is extended with \emph{potential functions} that track the resources associated with an expression.
\citet{ngo_bounded_2018} observed that this idea could be adapted to reasoning about expected resource consumption in randomized programs by taking weighted averages of potential functions when randomized choices are made, much as \ruleref{ht-rand-exp} in \thetimelogic takes a weighted average of credits. 
They develop a static analysis tool called Absynth that automatically derives expected resource bounds for first-order randomized programs.
The soundness of Absynth is justified in terms of derivations in a Hoare-like logic with judgements of the form $\hoare{\Gamma_1; Q_1}{e}{\Gamma_2; Q_2}$, where the pre- and post-conditions are divided up into a logical predicate $\Gamma_i$, and a quantitative potential function $Q_i$.
In contrast, \thetimelogic represents resources as credit assertions, which can appear anywhere within specifications, allowing for stored credit assertions that depend on data structure invariants, such as in the hash table.
\citet{DBLP:journals/pacmpl/WangKH20} subsequently developed pRaML, an AARA-based type system for expected costs of higher-order probabilistic programs.
However, pRaML does not support mutable references, so examples combining higher-order functions and references like the meldable heaps in \cref{sec:randomized-heaps} are beyond its scope.

In work done concurrently with ours, \citet{lohse2024iris} developed \emph{ExpIris}, a variant of Iris that supports establishing bounds on the expected cost of higher-order programs with mutable state.
ExpIris adds additional parameters to the standard weakest precondition of Iris \cite{irisjournal} that record an initial potential \(p : \real\) and a final potential \(\mathfrak{P} : \Val \ra \real\). The cost of a program is defined via manually inserted \texttt{tick} operations, where each \texttt{tick}(\(c\)) consumes \(c\) from the initial potential.
Because ExpIris tracks potentials through weakest precondition parameters, there are some restrictions on the use of potentials that are similar to those in the Hoare logic of \citet{ngo_bounded_2018} described above.
In particular, the final potential can only depend on the return value of a program but not on its final state.
\thetimelogic{} can handle such cases by breaking the tight coupling between the weakest precondition and costs through the use of the credit resource.\footnote{\ifbool{fullversion}{We prove the example discussed as ``out of scope'' in \citet[\S7]{lohse2024iris} in \appref{app:lohse-garg-example}.}
{The example discussed as ``out of scope'' in \citet[\S7]{lohse2024iris} is similar to a single invocation of \(\amortizedop\) as defined in \cref{sec:repeated-amortized-operations} and has a simple proof in \thetimelogic.}}
As an orthogonal consideration, ExpIris also supports reasoning about the total \emph{work} of concurrent programs (but not the \emph{span}, in the sense of complexity of parallel algorithms).

Analyses and logics based on Ranking Supermartingales (RSMs) have been developed for proving almost-sure termination, deriving high-probability bounds on random variables, bounding expected values and higher moments of running time, \eg, its variance~\cite{DBLP:conf/popl/FioritiH15, fu_termination_2019, McIverMKK18, DBLP:journals/pacmpl/AgrawalC018, chatterjee_quantitative_2024,kura_tail_2019}.
An RSM can be seen as a function assinging an expected termination time to every line in a program, depending on the state.
These works focus on supporting automation for first-order programs; RSMs are inferred using template-based synthesis and proven using arithmetic solvers.
To enable this, assignments and conditional guards are often required to be written in a particularly well-behaved subset of expressions, \eg, polynomials.
It would be interesting to investigate the idea that a \thetimelogic proof of a cost bound implicitly ensures the existence of a certain RSM. %

Some other approaches provide new formalisms to reason about expected runtime of
higher-order programs. \citet{avanzini_type_based} introduce a probabilistic extension
of sized types. This is a rich type system, which includes refinement types as well
as distributions types which can be thought of as probabilistic counterparts of sum types.
Other type-based approaches to
reason about almost-sure termination of probabilistic programs include intersection
types~\cite{dal_lago_intersection_2021}. Another approach by~\citet{cps} instead proposes a
continuation-passing style translation of a program into a runtime transformer, \ie,
a translated program that computes the runtime of the original program. Then, one
can use a standard higher-order program logic to reason about the translated program
and obtain runtime bounds on the original program.

\paragraph{Separation logics for probabilistic programs}

In addition to QSL and its variants, several other separation logics have been developed for reasoning about probabilistic programs.
Starting with PSL~\citep{DBLP:journals/pacmpl/BartheHL20}, several separation logics have been developed in which the notion of separation encoded in the separating conjunction corresponds to probabilistic independence~\citep{DBLP:journals/pacmpl/BaoGHT22,DBLP:journals/pacmpl/Li0H23,DBLP:conf/lics/BaoDH021}. 
Others re-use a standard interpretation of separating conjunction, but encode a probabilistic property in the interpretation of Hoare triples or weakest preconditions.
For example, Polaris~\citep{DBLP:journals/pacmpl/TassarottiH19} and Clutch~\citep{clutch} are Iris-based logics for doing probabilistic relational reasoning using the technique of probabilistic couplings~\citep{DBLP:conf/lpar/BartheEGHSS15, lindvall_lectures_2002}.
Caliper \citep{caliper} exploits probabilistic couplings to show almost-sure termination through termination-preserving refinement of a Markov chain model.
Eris~\citep{eris} is a variant of Iris that adapts the ideas of approximate Hoare Logic~\citep{DBLP:conf/icalp/BartheGGHS16} to enable proving that a specification holds with high probability.
Eris uses assertions called \emph{error credits}, where an error credit of value $\varepsilon$ can be ``spent'' to exclude reasoning about a randomized outcome that occurs with probability $\varepsilon$.
These error credits can be averaged in an expectation-preserving way with a rule similar to \ruleref{ht-rand-exp}.

\paragraph{Separation logics for non-probabilistic cost analysis}

The idea of using a separation logic resource to account for the runtime of a program
is due to~\citet{DBLP:journals/corr/abs-1104-1998}. Time credits were introduced to Iris by~\citet{tciris} in a logic called Iris$^{\$}$.
Besides time credits, which represent upper bounds on the time before a good event happens,
they also introduce a dual notion of time receipt, which allows lower bounding the time before a bad event happens.
In later work, \citet{pottier_thunks_2024} show how to encode Okasaki's debit-based reasoning~\citep{DBLP:books/daglib/0097014} about thunks in Iris$^{\$}$.
When reasoning about non-probabilistic parts of a program, the time credits in \thetimelogic have equivalent
reasoning rules as cost credits in Iris$^{\$}$.
Thus, the proofs of sophisticated examples developed in these previous works could also be carried out in
\thetimelogic.

However, it would not have been possible to extend the elegant techniques of \tciris to support probabilistic reasoning:
In \tciris deterministic time credits are realized by (1) translating the program to an instrumented program that has a distinguished counter that decreases at every step, (2) making the program crash if this counter reaches~0, and (3) making the time credits track the value of this counter.
At this point, the standard Iris adequacy theorem \cite{irisjournal} that ensures crash-freedom also enforces that this counter can never reach~0.
The program thus cannot take more steps than the initial amount of credits.
But this method cannot be applied to establish \emph{expected} cost bounds because one needs to increase the amount of credits in some execution branches and decrease in others. This would require reasoning about the expected value of a program location, and this is not supported in any existing logic for higher-order randomized stateful programs.

We can summarize our solution as follows.
First, we define a new resource algebra representing expected cost as well as a new bespoke weakest precondition that tracks this resource algebra and connects it to the probabilistic operational semantics of \thelang.
This weakest precondition supports distributing cost credits across execution branches via the expected cost modality.
We then prove that this weakest precondition interacts with step indexing and other Iris features such as invariants. %
Support for step indexing and handling this interaction is necessary to be able to recover all of the expressive features of Iris, such as reasoning about higher-order state.
Next, we prove all of the "standard" program logic rules which one may expect from, \eg, \citet{irisjournal}, as well as the novel cost-averaging rules.
Finally, we prove an adequacy theorem that connects a Hoare triple in \thetimelogic to the expected cost of a program.
The proof of the adequacy theorem intricate since it requires approximating the expected cost inductively, and connecting to the expected cost of full-program execution via a limiting argument.

\subsection{Future Work}
\label{sec:future-work}

We see several interesting research directions for extending \thetimelogic.

 \paragraph{Continuous distributions}
 \label{sec:continuous}

Adding support for sampling from continuous distributions to \thelang will bring significant challenges in terms of the mechanization of the measure theory needed to construct the operational semantics of our language and its associated metatheory.
However, these challenges are orthogonal to cost analysis, and we believe a similar rule as the expectation-preserving composition for cost credits should hold in the continuous setting.

 \paragraph{Cost lower bounds}
 \label{sec:lower-bounds}

 Some program logics for cost analysis support reasoning about lower bounds on costs.
 In the non-probabilistic setting, this can be achieved through \emph{time receipts}~\cite{tciris}, but for reasons analogous to the discussion of time credits in \cref{sec:related-work}, the technique used by \citeauthor{tciris} of deriving cost bounds via a program translation cannot directly be transposed into the probabilistic setting.
We expect that such an extension would require adding native support for time receipts at the level of the logic and hence the model of \thetimelogic.
However, extending \thetimelogic with support for lower bounds would be a challenge since, in general, proving lower bounds via an inductive argument is not sound and one has to impose additional conditions~\cite{lowerbound}. %
Alternatively, a \emph{linear} separation logic could allow dualizing the credit-style representation of upper bounds as a \emph{prerequisite} for taking steps into an \emph{obligation} to take steps. %

\section{Conclusion}
\label{sec:conclusion}
We presented \thetimelogic, the first program logic for proving upper bounds of the expected cost of higher-order probabilistic programs with local state. We developed the notion of probabilistic cost credits, derived from time credits that were originally used to prove the runtime cost of deterministic programs, which enables us to reason about the cost of \emph{probabilistic} programs.  Our definition of cost credits can depend flexibly on user-defined cost models, enabling reasoning about a much richer class of costs specific for different applications.  We demonstrated the strength of \thetimelogic on various examples, including ones whose runtime bounds are non-trivial or are amortized, which to our knowledge are outside the scope of prior techniques.

\begin{acks}
  We thank Alexandre Moine for his close reading and useful suggestions.
  This work was supported in part by the \grantsponsor{NSF}{National Science Foundation}{}, grant no.~\grantnum{NSF}{2225441}, the \grantsponsor{Carlsberg Foundation}{Carlsberg Foundation}{}, grant no.~\grantnum{Carlsberg Foundation}{CF23-0791}, a \grantsponsor{Villum}{Villum}{} Investigator grant, no. \grantnum{Villum}{25804}, Center for Basic Research in Program Verification (CPV), from the VILLUM Foundation, and the European Union (\grantsponsor{ERC}{ERC}{}, CHORDS, \grantnum{ERC}{101096090}).
  Views and opinions expressed are however those of the author(s) only and do not necessarily reflect those of the European Union or the European Research Council.
  Neither the European Union nor the granting authority can be held responsible for them.
\end{acks}

\section*{Data-Availability Statement}

A full Coq formalization of \thetimelogic including the case studies is available~\cite{tachis-artefact}.

\nocite{uniform_hash_assumption}
\bibliography{refs}

\ifbool{fullversion}{
\pagebreak
\appendix

\section{Full Definition of the Weakest Precondition}
\label{sec:app-wp}

The weakest precondition that \thetimelogic is based on is defined by the following guarded recursive fixed point.
\begin{align*}
  \wpre{\expr_1}[\mask]{\Phi}
  \eqdef{}
  &(\expr_1 \in \Val \mathrel{\land} \pvs[\mask] \Phi(\expr_1))\\
  \lor{}\,
  &(\expr_1 \not\in \Val \mathrel{\land} \All \state_1, \rt_1 . \stateinterp(\state_1) \sep \tcauth{\rt_1} \wand \pvs[\mask][\emptyset]\\
  & \quad
    \ecm
    (\expr_1, \state_1, \rt_1,
    \ (\Lam \, \expr_2, \state_2, \rt_2
    \ . \later \pvs[\emptyset][\mask] (\stateinterp(\state_2) \sep \tcauth{\rt_2} \sep \wpre{\expr_2}[\mask]{\Phi})))
\end{align*}
It differs from the presentation in \cref{sec:wp} in the use of the Iris update modality $\pvs$, which we earlier omitted for the sake of readability. The purpose of the update modality is to allow the user of the logic to perform updates to the resources they own; see, \eg, \citet{irisjournal} for details.

\section{Hash Map}
\label{sec:hash-map-appendix}

In this example, we establish bounds on inserting and looking-up elements in a hash map. We use the $\costtick$ model, where we incur $\tc{1}$ whenever we perform an access to memory by manually adding a $\Tick 1$ operation after each dereference.

Recall from \cref{sec:hash-map} that a hash map supports two operations: insertion and lookup. The hash map contains an array, where each array index contains a pointer to a linked list (called a bucket), which stores a list of elements. To insert an element, the hash map uses a hash function to hash the element, and append it to the bucket with the array index of the hash. On the other hand, to perform a lookup on an element, the hash map hashes the element, and traverse the bucket with the array index of the hash to determine whether it is present in the bucket.

We first implement a model of the idealized hash function in \cref{fig:app_hashfun} under the uniform hash assumption~\cite{uniform_hash_assumption}, which assumes that the hash function $hf$ from a set of keys $K$ to values $V$  behaves as if, for each key $k$, the hash $hf(k)$ is randomly sampled from a uniform distribution over $V$, independently of all other keys.
We implement this model by first initializing an empty mutable map $hf$. When we want to hash a key $k$, we first check the map to determine whether there is an element under that key.
If so, we return the hash value stored in $hf(k)$.
Otherwise, we sample a value uniformly from $V=\{0, \dots, n\}$, store the value in $hf$ with key $k$, and return it.

\captionsetup{belowskip=-5pt,aboveskip=5pt}

\begin{figure*}[htb]
  \centering
  \begin{align*}
    \computehash \eqdef{}
    & \Lam\ hf\ k.\MatchML \mapget\ hf\ k with
      \Some(v) => v
      | \None =>
      {\begin{array}[t]{l}
         \Let v = \Rand n  in \\
         \mapset\ hf\ k\ v;    \\
         v
       \end{array}}
      end {}
  \end{align*}  
  \caption{An idealized hash function implementation.}
  \label{fig:app_hashfun}
\end{figure*}

Note that in this example, although the implementation of maps involve various location accesses, we do not include $\Tick$ operations in our hash implementation.
This is because our implementation only serves to simulate the effects of a realistic hash function, which in real life, a hash function usually incurs constant  cost for each hash operation.

We then implement the hash map as a client of the hash function. We represent our hash map $h$ as a tuple $(l, hf)$. The former is an array whose elements are buckets of the hash map while the latter is the hash function. We represent the list of elements which we store in the bucket with a linked list.

The hash map supports two main operations shown in \cref{fig:app_hashmap}: $\insertelem$ which inserts an element into the hash table, and $\lookupelem$ which determines whether an element is stored in it.
The function $\insertelem$ hashes an element with the hash function $hf$ to get the index of the bucket it should go to.
Then we access that bucket with a dereference, incurring one tick.
Finally we insert the element into the end of the linked list representing the bucket.
Here, when we traverse to the end of the linked list, we also perform a check to determine whether the element already exists in the bucket, in which case, we do not add it at the end of the linked list to avoid duplication.
The function $\lookupelem$ works similarly by first hashing the element with $hf$ to get the index of the bucket, accessing the bucket with a load from memory and hence incurring one tick cost, and finally performing a lookup by traversing the linked list.
Since both linked list functions $\llinsert$ and $\lllookup$ traverses the entire linked list, and memory dereferencing costs $1$ in our model, the functions incur $\tc{l}$ credits for each invocation where $l$ is the length of the linked list.

\begin{figure*}[t]
  \centering
  \begin{minipage}[t]{.4\linewidth}
    \begin{align*}
      \insertelem \eqdef{} & \Lam\ h\ v.\\
                           &\Let (l, hf) = h in \\
                           & \Let idx =\computehash\ hf\ v in \\
                           & \Let b = \deref l[idx] in \\
                           & \Tick{1}; \\
                           &(l\locadd idx) \gets \llinsert\ b\ v\\
    \end{align*}
  \end{minipage}
  \begin{minipage}[t]{.4\linewidth}
    \begin{align*}
      \lookupelem \eqdef{} & \Lam\ h\ v.\\
                           &\Let (l, hf) = h in \\
                           & \Let idx =\computehash\ hf\ v in \\
                           & \Let b = \deref l[idx] in \\
                           & \Tick{1}; \\
                           &\lllookup\ b\ v
    \end{align*}
  \end{minipage}
  \caption{A hash map implementation.}
  \label{fig:app_hashmap}
\end{figure*}

Let us consider the  costs of $\insertelem$ and $\lookupelem$ in the case where an element has never been hashed by a hash function $hf$ before. For $\insertelem$, when we first hash the element, since it has not been hashed by $hf$ before, we randomly pick an index $idx$ from $V=\{0, \dots, n\}$. Let the  cost of $\insertelem$ in the case that the element is hashed to $idx$ by $hf$ be $C(idx)$. By stepping through the program, we can show that $C(idx) = 1 + \hmlength(h, idx)$, where $\hmlength(h, idx)$ is notation for the length of the linked list representing the $idx$-th bucket. The first $1$ cost comes from accessing the $idx$-th bucket, and $\hmlength$ comes from inserting the element at the end of the linked list. Hence the expected  cost of $\insertelem$ is $\expect[\mathbb{U}(0, n)]{C}$, where $\mathbb{U}(0,n)$ is the discrete uniform distribution from $0$ to $n$. By linearity of expectations, we have $\expect[\mathbb{U}(0, n)]{C}=\expect[\mathbb{U}(0, n)]{1}+\expect[\mathbb{U}(0, n)]{\hmlength(h, idx)}=1+\frac{\hmsize(h)}{n+1}$ where $\hmsize(h)$ is shorthand for the total number of elements stored in the hash map (also the sum of lengths of all the linked list buckets). The time complexity of $\lookupelem$ is exactly the same, as instead of adding an element to the end of a linked list in $\insertelem$, we still traverse the linked list comparing our element with each value stored by the linked list.

To prove this claim, we use \thetimelogic to prove the following two specifications.
\begin{align*}
  &\hoareV
    {\val \notin \dom(m_{1}) \sep
    \ishashmap(h, m_1, m_2) \sep 
    \tc{1+\tfrac{\hmsize(h)}{n+1}}
    }
    {\insertelem~h~\val}
    {\_ \ldotp \Exists i . \ishashmap(h, \mapinsert{\val}{i}{m_1}, \mapinsert{i}{m_2[i] \dplus [\val]}{m_2})} \\
  &\hoareV
    {
    v\notin \dom(m_1) \sep 
    \ishashmap(h, m_1, m_2) \sep 
    \tc{1+\tfrac{\hmsize(h)}{n+1}}
    }
  {\lookupelem\ h\ v}
  {b \ldotp b = \False \sep \Exists i . \ishashmap(h, \mapinsert{v}{i}{m_1}, m_2)}
\end{align*}
Intuitively, the predicate $\ishashmap\ h\ m_1\ m_2$ asserts that the hash map $h$ is valid according the two mathematical maps $m_1$ and $m_2$. For the rest of the subsection, it suffices to understand that the abstract predicate contains two main components, that the abstract map $m_1$ tracks the concrete map implementing the hash function, which we represent with the predicate $\ishashfunction$, and the abstract map $m_2$ tracks the linked lists of each bucket of the concrete hash map. Consequently, the proposition $v\notin \dom(m_1)$ encodes the idea that the hash function has not hashed the element $v$ before.

The proof of the two above specifications are fairly standard, except for the part where we hash the element $v$. Note that we have to split our cost credits $1+\frac{\hmsize(h)}{n+1}$ somewhere but at first glance there does not seem to be any $\Rand$ expressions for us to apply the expectation rule. The reason is that the $\Rand$ expression is encapsulated in the hash function and is not exposed for the client. This subtle detail means that we have to provide a strong specification of the hash function that allows us to split our credits, just as we could for a normal $\Rand$ expression.

As an example, the following specification for the hash function is not strong enough for verifying the hash table albeit being valid:
\begin{align*}
  \hoare
  {v\notin \dom(m_1)\sep \ishashfunction\ hf\ m_1}
  {hf\ v}
  {idx \ldotp \ishashfunction\ hf\ \mapinsert{v}{idx}{m_1}}
\end{align*}
The reason is that the specification does not allow us to split credits in any meaningful way. To prove the specification of the hash map functions, one have to use the following strictly stronger specification for the hash function which allows the client to divide the credits depending on the hash value, similarly to the expectation rule for $\Rand$:
\begin{align*}
  \hoareV
  {
  v\notin \dom(m_1)\sep
  \ishashfunction\ hf\ m_1\sep 
  \Sigma_{i=0}^n (\tfrac{x_2(i)}{n+1})=x_1\sep
  \forall i, 0\le x_2(i) \sep 
  \tc{x_1}
  }
  {hf\ v}
  {idx \ldotp \ishashfunction\ hf\ \mapinsert{v}{idx}{m_1}\sep \mhl{\tc{x_2(idx)}}}
\end{align*}

One limitation of the specification of $\insertelem$ is that required credits for each insertion is propositional to the size of the hash table, which leads to worse modularity as a client of the hash table needs to know the exact size of the hash table. We can alleviate this problem by providing an amortized specification of $\insertelem$ where the cost credits required for each insertion is constant. The tradeoff is that one has to fix the total maximum size of the hash table \textit{a priori}. If the total maximum size of the hash table is $\hmmax$, the total number of  cost is hence $\tc{\Sigma_{i=0}^{\hmmax-1} (1+\frac{\hmsize(h)}{n+1})}= \hmmax + \frac{(\hmmax-1)\cdot \hmmax}{2(n+1)}$. If we average out the credits, the average  cost is simply $\tc{\frac{\hmmax}{\hmmax} + \frac{(\hmmax-1)\cdot \hmmax}{2(n+1)\cdot \hmmax}}= \tc{1+\frac{\hmmax-1}{2(n+1)}}$, which we simply denote as $\tc{x}$. Therefore we can write the following amortized spec:
\begin{align*}
  \hoareV
  {v\notin \dom(m_1) \sep 
  \hmsize(h)<\hmmax \sep 
  \isamortizedhashmap(h, m_1, m_2) \sep 
  \tc{x}}
  {\insertelem~h~v}
  {\_ \ldotp \Exists i . \isamortizedhashmap(h, \mapinsert{v}{i}{m_1}, \mapinsert{}{m_2[i] \dplus [v]}{m_2})}
\end{align*}
Here we make three main changes to the original specification of $\insertelem$. First we require that the total size of the hashmap is smaller than $\hmmax$. We also fix the credit in the precondition to a constant. Lastly, we change the abstract predicate $\ishashmap$ to $\isamortizedhashmap$. Intuitively, $\isamortizedhashmap$ not only contains the $\ishashmap$ abstract predicate but also the extra time credits one pays in excess for the first half of the insertions. For the last half of the insertions, the amortized cost is insufficient to pay for the expensive operation so we take just enough from this reserve to pay for the difference. To highlight the modularity aspect of \thetimelogic, we actually derive this amortized specification from the non-amortized one of the $\insertelem$ function.

We note that we are unable to provide a meaningful amortized specification for the $\lookupelem$ function as each invocation of the function does not change the size of the hash table, and hence we are unable to fix the number of times we can call $\lookupelem$.

We also note that in \thetimelogic, we are unable to provide a  cost that is proportional to the size of the hash table for the case where a key has been hashed before.
In textbook proofs, one can argue that the cost is still proportional to the size of the hash table even if the key has been hashed before (or inserted into the hash table) assuming that
the element is chosen uniformly across all elements stored in the hash table. However, our logic cannot describe this assumption and we leave this extension to our logic as future work.

\section{State-dependent Expected Cost}
\label{app:lohse-garg-example}

We prove that the following program, suggested by \cite[\S 7]{lohse2024iris}, has expected cost~\(\frac 1 2\).
\begin{align*}
  \langv{toss}~\langv{l} \eqdef{} & \If \Rand 1 = 1 then \TT \\
                              & \Else \langv{l} \gets \deref \langv{l} + 1 \\[1mm]
  \langv{T}_{\!\frac 1 2} \eqdef{} & \Let \langv{l} = \Alloc 0 in \langv{toss}~\langv{l} \Seq \Tick(\deref \langv{l})
\end{align*}
According to \loccit, the analysis of the expected cost of \(\langv{T}_{\!\frac 1 2}\) is interesting in that the cost depends on the value stored in the location \(\langv{l}\).

We can derive the cost bound in \eclog by choosing the \(\costtick\) model. To begin, we allocate a reference \(\ell\) and perform an expectation-preserving splitting of the cost credits \(\tc{\frac 1 2}\). We give \(\tc{f(1)} = \tc 0\) to the case where \(\Rand 1\) returns \(1\) and \(\ell\) is not incremented, and reserve \(\tc{f(0)}=\tc 1\) for the \(\langkw{else}\) branch.

\begin{gather*}\footnotesize
  \inferrule*[Right=bind]{
    \infrule*[Right]{rand-exp} {f(n) \eqdef \text{if } n=1 \text{ then } 0 \text{ else } 1
      \and
      0 + \expect f \leq \textstyle \frac 1 2 }
    {\inferrule*[right=frame]{\hoare {\tc{\textstyle \frac 1 2}} {\Rand 1}
        {n\,.\,\tc{f(n)}}}
      {\hoare {\ell \mapsto 0 \sep \tc{\textstyle \frac 1 2}} {\Rand 1}
    {n\,.\,\ell \mapsto 0 \sep \tc{f(n)}}}}
    \and
    \inferrule*{ (1) }
    {\All n .
      \hoare {\ell \mapsto 0 \sep \tc{f(n)}}
      {\langv{toss}~\ell \Seq \Tick(\deref \ell)}
      \TRUE}
  }
  {
    \inferrule*
    {\hoare {\ell \mapsto 0 \sep \tc{\textstyle \frac 1 2}} {\langv{toss}~\ell \Seq \Tick(\deref \ell)} \TRUE}
    {\hoare {\tc {\textstyle \frac 1 2}} {\langv{T}_{\!\frac 1 2}} \TRUE}
  }
\end{gather*}

The result now follows from a straightforward case analysis and the definition of \(\costtick\).

\begin{gather*}
  \footnotesize
  \inferrule*[left={case \(n=1\)\,?}]{
    \inferrule*[Left={\footnotesize{load,frame}}]{
      \inferrule*[Left=tick]{ }
      {\hoare {\tc 0} {\TT \Seq  \Tick(0)} \TRUE}}
    {\hoare {\ell \mapsto 0 \sep \tc{f(1)}} {\TT \Seq  \Tick(\deref \ell)} \TRUE}
    \and
    \inferrule*[right={\footnotesize{load,store}}]{
      {
        \inferrule*[Right={\footnotesize{load,frame}}]
        {\inferrule*[Right=tick]{ }
          {\hoare{\tc{f(1)}}
            {\Tick(1)} \TRUE}}
        {\hoare{\ell \mapsto 1 \sep \tc{f(1)}}
          {\Tick(\deref \ell)} \TRUE}}
    }
    {\hoare{\ell \mapsto 0 \sep \tc{f(0)}}
      {\ell \gets \deref \ell + 1 \Seq \Tick(\deref \ell)} \TRUE}
  }
  {\inferrule*{\All n .
    \hoare {\ell \mapsto 0 \sep \tc{f(n)}}
    {(\If n = 1 then \TT \Else \ell \gets \deref \ell + 1) \Seq \Tick(\deref \ell)}
    \TRUE}{(1)}}
\end{gather*}

The expected cost bound then follows directly by adequacy.

\vfill

}{}

\end{document}